%% file: main.tex
\documentclass[english,journal,comsoc]{IEEEtran}
\usepackage{algorithm}
\usepackage{algorithmicx}
\usepackage{algpseudocode}
\usepackage{array}
\usepackage{subcaption}
\usepackage{textcomp}
\usepackage{stfloats}
\usepackage{url}
\usepackage{verbatim}
\usepackage{graphicx}
\usepackage[doi=false, url=false, isbn=false, eprint=false, style=ieee, backend=biber]{biblatex}
\usepackage{xcolor}
\usepackage{babel,csquotes}
\input{header.tex}
\algrenewcommand\algorithmicrequire{\textbf{Input:}}
\algrenewcommand\algorithmicensure{\textbf{Output:}}

\begin{document}

\title{Goal-oriented Resource Allocation for \\ Collaborative Integrated Sensing and Communication}

\author{
\IEEEauthorblockN{Tran Trong Duy, Maxime Ferreira Da Costa, Salah Eddine Elayoubi, and Nguyen Linh Trung}

% \thanks{Manuscript received xxx; revised yyy.}
\thanks{This work has been partially supported by the Sustainable 6G chair, funded by Orange and held by CentraleSupélec. It was also supported by the French government, in the framework of the France Excellence Vietnam Scholarship Program and the France 2023 program (INTENTION-6G project). Tran Trong Duy and Nguyen Linh Trung were funded by the research
project QG.25.08 of Vietnam National University, Hanoi.}% <-this % stops a space
%\thanks{Tran Trong Duy is with Laboratory of Signals and Systems, CentraleSup\'elec, Universit\'e Paris-Saclay, CNRS, Gif-sur-Yvette, France, and also with University of Engineering and Technology, Vietnam National University, Hanoi, Vietnam (e-mail: trong-duy.tran@centralesupelec.fr).}
\thanks{Tran Trong Duy, Maxime Ferreira Da Costa and Salah Eddine Elayoubi are with Laboratory of Signals and Systems, CentraleSup\'elec, Universit\'e Paris-Saclay, CNRS, Gif-sur-Yvette, France (e-mail: \{trong-duy.tran; maxime.ferreira; salaheddine.elayoubi\}@centralesupelec.fr).}
\thanks{Nguyen Linh Trung is with University of Engineering and Technology, Vietnam National University, Hanoi, Vietnam (e-mail: linhtrung@vnu.edu.vn).}
}

% \markboth{IEEE Transactions,~Vol.~x, No.~y, z~2026}%


\maketitle

\begin{abstract}
In this paper, we consider resource allocation for a collaborative integrated sensing and communication (ISAC) scenario, in which distributed smart devices can be scheduled to perform sensing and transmit their sensing features to a fusion center. The fusion center aims to perform classification tasks on the environment based on received features. A scalable network-sensing framework is proposed to balance the performance of the sensing service with that of the classical enhanced Mobile Broadband (eMBB) service. We adopt a tractable theoretical metric, the discriminant gain, as a proxy for the classification goal. We formulate cross-layer optimization problems to maximize discriminant gain under constraints on energy consumption and eMBB communication quality for the independent and joint scheduling policies. The joint scheduling policy has considerably higher complexity than the independent scheduling policy, in exchange for better collaborative sensing performance. A simplified gain model is proposed to reduce the complexity and practicality of the joint scheduling policy. Both policies are obtained via successive convex approximation and parametric convex optimization. Extensive experiments are conducted to verify the goal-oriented framework and the two policies. It is demonstrated that the two policies outperform the baseline policies with both synthetic and realistic radar simulation datasets. The joint scheduling policy can exploit device correlations and thus performs better than the independent scheduling policy under strong correlations and strict communication constraints.
\end{abstract}


\section{Introduction}
Integrated Sensing and Communication (ISAC) integrates radar sensing and wireless communications into a single hardware platform and uses RF signals for sensing \cite{liu2022integrated}. As one of the key use cases for future sixth-generation (6G), ISAC has been extensively studied over the past few years \cite{Zhang_2026_ISAC_OTY}. 
While the majority of ISAC literature focuses on the performance trade-offs between sensing and communications for an optimal physical-layer design \cite{chalise2017performance,liu2021cramer,guerci2015joint}, there remains a need for a global design framework that adopts a holistic view of networking and sensing constraints and goals. 

We argue in this paper that goal-oriented communication is an adequate paradigm in this context. Goal-oriented communication has emerged as a novel paradigm in wireless networks, and is expected to be an integral part of 6G systems \cite{BeyondShannon6GNetworks,getu2024survey}. The goal relates to the performance of a specific application rather than classical Quality of Service (QoS) metrics (throughput, delay, etc.). In the specific context of ISAC, the goal relates to sensing performance and should be integrated into the system's overall design. 

In this paper, we consider a collaborative ISAC system and adopt the discriminant gain \cite{Lan_2023_ProgressiveFTX} to measure sensing performance, while the classical communication requirement, measured in terms of data rate for the enhanced Mobile Broadband (eMBB), is considered as a constraint~\cite{series2015imt}. We investigate the following research question: \emph{How to allocate resources to enhance collaborative sensing performance while maintaining a target QoS level for classical communication?} 

Our goal-oriented design leverages two factors: sensing power and scheduling policy. Two scheduling policies, called \emph{independent} and \emph{joint} are proposed in Sections~\ref{sec:independent_policy} and~\ref{sec:joint_policy}, respectively. The independent scheduling policy assumes that devices' contributions to the common goal are independent; as a result, the scheduling problem is significantly simplified. In contrast, the joint scheduling policy accounts for performance improvements enabled by the cooperation of every device pair. This enables better collaborative sensing, but at the cost of more complex scheduling. The joint scheduling policy theoretically requires knowledge of the correlation matrix between sensing features from different devices to compute the joint sensing gain. A simplified joint gain model is proposed to ease the optimization of scheduling parameters and improve the practicality of the policy by avoiding the need to estimate a large correlation matrix. Both policies aim to maximize sensing accuracy while satisfying constraints on device power, eMBB data rate, and total system energy.

The original contributions of the paper are as follows:
\begin{itemize}
    \item We propose a novel goal-oriented system design framework for collaborative ISAC considering eMBB service quality, sensing performance, and energy consumption. Our cross-layer scheme thus integrates the physical, MAC (Medium Access Control), and application layers.
    
    \item We formulate a constrained optimization problem that ensures scalability by selecting the optimal subset of sensors to query within available resources, while maximizing the application goal. In particular, when there is correlation between devices, we propose a joint scheduling policy that can be optimized using successive convex approximation (SCA) and a sum-of-ratios algorithm.
    
    \item We validate the proposed policies using synthetic and simulation datasets and compare them to different baselines. The simulation datasets for independent and joint scheduling are recorded for human recognition using a frequency-modulated continuous-wave (FMCW) radar and for cooperative object detection using a 4D radar, respectively. The results demonstrate that the proposed scheme preserves sensing accuracy while meeting high-level guarantees for eMBB performance.
    
\end{itemize}

This work is an extension of our conference paper \cite{duy_GLOBECOM25}, relaxing the assumptions about the independence between data generated by different devices and the negligible impact of feature transmissions. This paper presents a complete analysis of the collaborative sensing framework in the presence of correlations. Additionally, it accounts for the cost of feature transmission, which is particularly important in complex applications, and presents extensive experiments. The remainder of this paper is organized as follows. Section \ref{sec:related} reviews the literature and identifies research gaps. In Sections \ref{sec:model} and~\ref{sec:design}, we describe the system model and introduce our goal-oriented design. Section \ref{sec:independent_policy} presents the independent scheduling policy as an optimization problem and provides tractable solutions. For the joint scheduling policy, we propose a simplified joint gain model in Section~\ref{sec:joint_policy} and discuss how to sample a schedule from it. Section~\ref{sec:experiments} gives details of numerical experiments and discusses the results. Finally, Section~\ref{sec:conclusion} concludes the paper with some future perspectives.

\begin{figure*}[t]
    \centering
    \includegraphics[width=0.9\linewidth, trim={0 0 0 90}, clip]{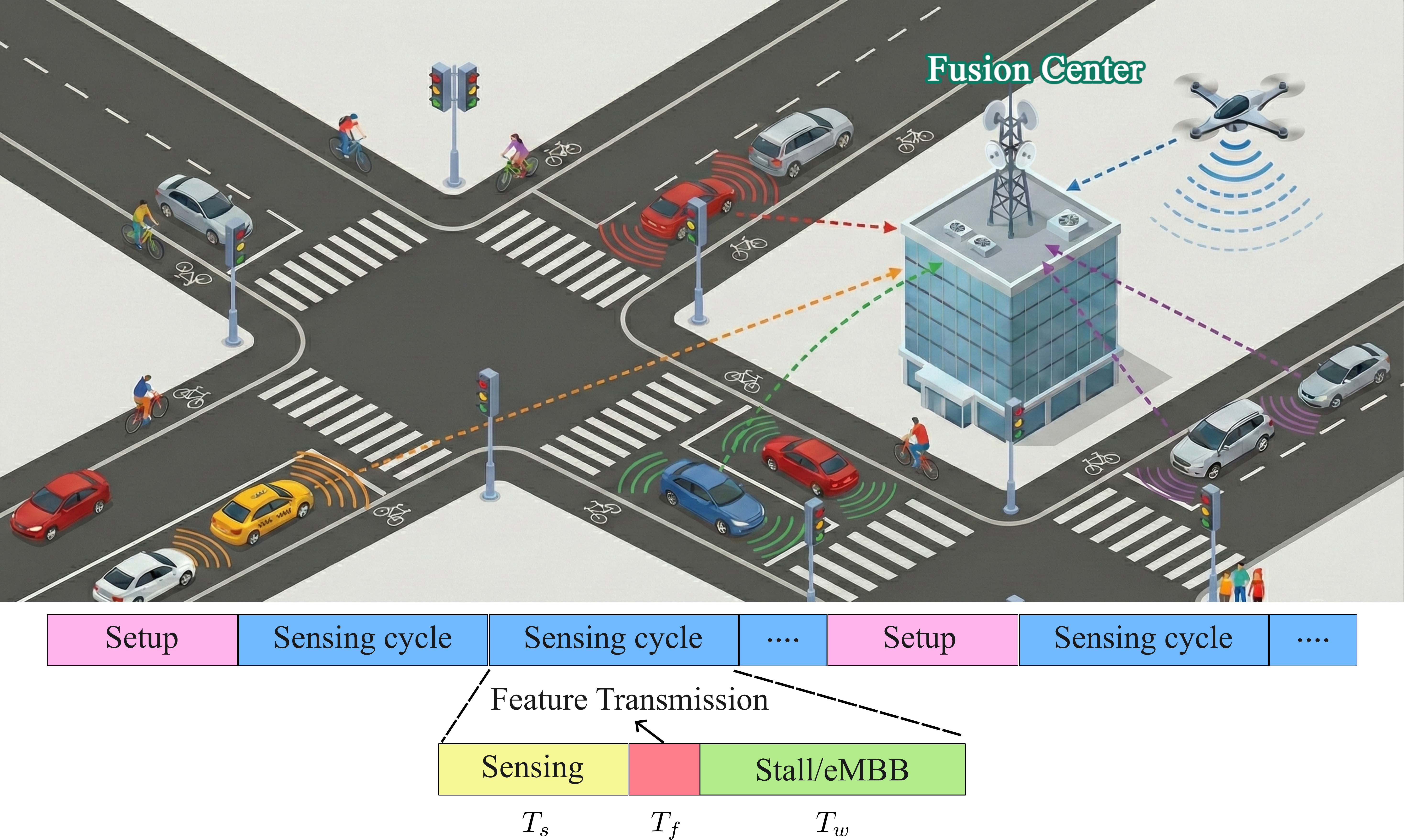}
    \caption{ISAC Network Sensing Model (Partially generated by Google Gemini).}
    \label{fig:system_model}
\end{figure*}

\section{Related works}\label{sec:related}

\subsection{ISAC}

Fundamental performance limits for ISAC have been studied from an information-theoretical perspective in \cite{liu2022survey}. From a physical-layer perspective, the trade-off between radar metrics (detection probability, resolution, etc.) and the communication capacity has been studied in \cite{chalise2017performance,liu2021cramer,guerci2015joint}. Major efforts are devoted to the physical aspects of ISAC, including beamforming, waveform design, channel modeling, and physical-layer security. 
 However, ISAC is not restricted to the physical layer and encompasses the broader concept of cross-layer design for 6G networks to convey sensing information in an optimal way \cite{liu2022integrated,Wen_2024_ISCC,da2024goal,Meng_Network_ISAC_2024}. A distributed source estimation problem is considered in~\cite{da2024goal}, where sensors have partial information about the source and use the communication channel to share data and perform estimation. \cite{Meng_Network_ISAC_2024} shows that considering network-level ISAC benefits both sensing and communication and presents several topologies for cooperative ISAC networks and their opportunities. One of the prominent research problems on ISAC is to develop ISAC-aware scheduling/resource allocation policies that dynamically balance spatial, temporal, and spectral resources \cite{Zhang_2026_ISAC_OTY}. Goal-oriented communication is a promising approach to address this problem.
 
\subsection{Goal-Oriented communication}
Goal-oriented communications aim to maximize the impact of the received bits towards a custom goal intended by the transmitter and the receiver \cite{BeyondShannon6GNetworks}\cite{gunduz2022beyond}.

The main difficulty in goal-oriented communication is measuring progress towards the goal from the application's perspective. A popular approach 
is to heuristically approximate the relationship between the goal's success and network variables by fitting an analytical function \cite{liu2023adaptable} or using a look-up table \cite{binucci2022adaptive}. However, these heuristics lack theoretical support and rely heavily on approximation errors in the data. From a theoretical perspective, \cite{Feng_2024_CPS} proposed to allocate resources based on the information utility gain, which is the difference between the system goal function before and after new data arrives. The main drawback is that the server needs to obtain information utility of new data before deciding which device will transmit. This non-causal problem creates overhead on the device side for utility transmission and assessment. 

Recently, in classification tasks, there has been growing interest in using a theoretical metric, called \emph{discriminant gain}, as a proxy for classification quality. The discriminant gain was first considered in \cite{Lan_2023_ProgressiveFTX} as a criterion for progressively selecting and transmitting features until a desired inference performance is reached, followed by multiple works that applied it in the wireless context \cite{Wen_2024_ISCC, Wen_2024_gain_OTA_MIMO_homo, Zhuang_2024_gain_SIMO_OFDM, Jiao_2024_gain_homo, Yang_2026_gain_MIMO, Chen_2024_Aggregation_Gain}. 

\subsection{Goal-oriented communications for ISAC}
The series of works by Wen et. al. \cite{Wen_2024_gain_OTA_MIMO_homo, Zhuang_2024_gain_SIMO_OFDM, Jiao_2024_gain_homo, Yang_2026_gain_MIMO} proposed a goal-oriented framework for over-the-air computation (AirComp), based on the discriminant gain. They mainly focused on precoding and beamforming designs that maximize the receiver's discriminant gain in the context of feature aggregation. Gain-based, goal-oriented AirComp has been extensively studied across different channel models, \emph{e.g.}, single-input-multiple-output (SIMO), orthogonal frequency-division multiplexing (OFDM), and multiple-input-multiple-output (MIMO). In \cite{Chen_2024_Aggregation_Gain}, the end-to-end (E2E) sensing performance analysis of feature aggregation was given as the number of sensors grows, showing that over-the-air computation is only beneficial when the degrees of freedom (DoFs) are insufficient for orthogonal access. The disadvantage of combining features over the air is that the number of features that can be aggregated is limited by the number of coherent timeslots, antennas, or OFDM subcarriers. Besides, AirComp feature aggregation requires all devices to deploy a homogeneous sensor and feature extraction scheme, as well as precise synchronization. 
The work of \cite{Wen_2024_ISCC} proposed a goal-oriented integrated sensing, computation, and communication (ISCC) scheme. The discriminant gain was optimized under constraints on time delay, feature transmission, and local energy for a sequential sensing and feature transmission paradigm. However, they assumed a fixed sensing waveform, a linear signal processing chain, and independent, non-overlapping sensing areas between devices. 

\subsection{Research gap and our positioning}

The benefit of cross-layer design and optimization for ISAC deserves further investigation \cite{Wymeersch_2025_cross_layer_ISAC} and we believe that goal-oriented design is a promising way for this aim. As the number of ISAC-capable devices is anticipated to rise, a predominant problem is how to manage resources to maximize their cooperative performance. While the framework in \cite{Wen_2024_ISCC} considers no cooperation between devices, other works \cite{Wen_2024_gain_OTA_MIMO_homo, Zhuang_2024_gain_SIMO_OFDM, Jiao_2024_gain_homo, Yang_2026_gain_MIMO, Chen_2024_Aggregation_Gain} show high potentials for AirComp cooperation, but with scalability problems. Additionally, to the best of our knowledge, none of the mentioned goal-oriented ISAC works consider the interplay between sensing and classical communication. In this work, the focus is on a collaborative ISAC system with a classification goal. We propose a goal-oriented ISAC framework that enables collaboration between devices, accounts for eMBB service performance, and integrates correlations across sensing data from different devices.

\section{System Model}\label{sec:model}

\subsection{Distributed Sensing Model}

We consider a scenario in which $K$ devices sense an environment in response to a query from a fusion center. The fusion center selects a policy that dictates the expected probability of selecting a device for the sensing task and the power a device must allocate to sensing. Then, at the beginning of a monitoring cycle, the devices selected for sensing according to the scheduling policy dedicate a share of their power to gather environmental data (\emph{e.g., radar sensing, video monitoring}) and to extract features from this recording. At the end of the monitoring cycle, the features are sent to the fusion center, which performs classification using the gathered features.

In this paper, we assume the fusion center's goal is to classify targets in the environment. To that end, the smart devices employ dual-functional radar and communication (DFRC) transceivers \cite{8386661} and only switch to sensing mode when scheduled. There are two phases in our goal-oriented system: setup and operating. In the setup phase, the fusion center decides on a policy that is then deployed in the subsequent operating phase; it may collect data from smart devices if necessary. The operating phase consists of multiple sensing cycles. After the operating phase ends, the fusion center re-evaluates the current state of the system, \emph{e.g.}, the number of connected smart devices, radio conditions, and device correlations, and designs a new policy to be deployed. A sensing cycle consists of three stages: sensing, feature transmission, and stall. In the sensing stage, the scheduled devices use radar signals to sense the environment. Then, they extract local features, transmit them to the fusion center, and finally stall until the start of the next sensing cycle. During this stalling period, their transceivers can be used for classical communication. When a device is not scheduled for sensing by the fusion center during the cycle, it uses its transceivers for classical communication throughout the entire cycle. Our system model is illustrated in Figure~\ref{fig:system_model}. In the following, we exclude the setup phase from the analyses as we do not specify a detailed implementation for it.

\subsection{Sensing Task and Network Classification Metric}
Let $T_s, T_f$, and $ T_{w}$ denote the duration of the sensing time, the feature transmission time, and the stalling (waiting) time (Figure~\ref{fig:system_model}), respectively. The duration of the sensing cycle $T$ is thus given by $T=T_s + T_f + T_w$.

After sensing, the $k$-th device extracts a feature vector $\bm{x}_k \in \mathbb{R}^{{N}_k}$ which we model as a random vector whose probability distribution depends on the state of the environment we want to estimate. We write $p_{k,\ell} = \mathbb{P}(\bm{x}_k | H_\ell)$ the conditional probability distribution of $\bm{x}_k$ given the hypothesis $H_\ell$ that the target belongs to the $\ell$-th class, $\ell \in \{1,2,\dots,L\}$.

For statistical and optimization convenience, we focus on the case of Gaussian conditional distributions of the classes:
\begin{equation}\label{eq:prob_distribution}
    p_{k,\ell} = \cN \left(\bm{\mu}_{k,\ell},  \bm{\Sigma}_{k} + \frac{1}{ P_{s,k}} \diag ( \bm{\eta}^2_{k}) \right)
\end{equation}
$\bm{\mu}_{k,\ell} = \{ \mu_{n_k,\ell} \}_{1_k}^{N_k}$ is the mean value of $\bm{x}_k$ under $H_\ell$. The residual variance of $\bm{x}_k$ and the noise variance are denoted $\bm{\Sigma}_{k}= \mathop{diag}(\sigma^2_{1_k}, \cdots, \sigma^2_{N_k})$ and $\bm{\eta}_k^2 = [\eta^2_{1_k}, \cdots, \eta^2_{N_k}]^T$, respectively, which are assumed to be independent of $\ell$, while $P_{s,k}$ is the power allocated by the $k$-th device to perform sensing. The variance of $p_{k,\ell}$ in~\eqref{eq:prob_distribution} decreases with $P_{s,k}$, and converges to the residual variance $\bm{\Sigma}_{k}$ in the asymptotic $P_{s,k} \to \infty$. %{\color{red} add support for GMM \cite[Chapter 8]{Couillet_Liao_RMT_ML}.}

We assess the goodness of the classification task in terms of its \emph{discriminant gain}~\cite{Lan_2023_ProgressiveFTX}, which measures the discernibility between the classes. At the $k$-th sensor, the pairwise discriminant gain between two hypotheses (or classes) $H_\ell$ and $H_{\ell^\prime}$ is defined as the symmetric Kullback-Leibler (KL) divergence between the conditional probability density $p_{k,\ell}$ and $p_{k,\ell^\prime}$. That is, under the Gaussian assumption~\eqref{eq:prob_distribution}:
\begin{align}
    G_k^{(\ell,\ell^\prime)}(P_{s,k}) &\triangleq \mathrm{D_{KL}}\left(p_{k,\ell} \,\Vert \, p_{k,\ell^\prime} \right) + \mathrm{D_{KL}}\left( p_{k,\ell^\prime} \,\Vert \, p_{k,\ell} \right) \nonumber \\ 
    &= \sum_{n_k=1}^{N_k}  \frac{(\mu_{n _k,\ell} - \mu_{n_k,\ell^\prime})^2}{\sigma^2_{n_k} + \eta^2_{n_k}/ P_{s,k}}.
    \label{eq:discriminant_gain}
\end{align}
The metric~\eqref{eq:discriminant_gain} is theoretically grounded since the KL divergence provides a lower bound of the error exponent of the probability of error of the optimal Hoeffding test~\cite{hoeffding1965asymptotically}.

In our distributed sensing model, the fusion center dictates a policy $\{\bm{\pi}, \bm{P}_s = [P_{s,1},\dots, P_{s,K}]\}$, where $\bpi$ controls the fraction of time that devices in the network have to allocate to the common sensing task with sensing power $P_{s,k}$. The pairwise network discriminant gain under a policy $\{\bm{\pi}, \bm{P}_s\}$ is defined at the fusion center as the linear combination of the sensor's discriminant gain
\begin{equation}\label{eq:G_pairwise_network}
    G^{(\ell,\ell^\prime)}(\bpi, \bP_{s}) =  g\left( \bpi, \bP_{s}, G_1^{(\ell,\ell^\prime)}, \dots, G_K^{(\ell,\ell^\prime)} \right),
\end{equation}
where $g(\cdot)$ is an aggregation function of device gains, and will be specified in Sections \ref{sec:independent_policy} and \ref{sec:joint_policy}.

Ideally, when the number of classes is greater than two, the \emph{network discriminant gain} is defined as the smallest discriminant gain between any two classes
\begin{equation}
\label{eq:Gtotal}
    \overline{G}(\bpi, \bP_s) = \min_{\ell\neq \ell^\prime} G^{(\ell,\ell^\prime)}(\bpi, \bP_{s})
\end{equation}
reflecting the least distinguishable pair of classes for the classification goal.
While the application goal is usually expressed in terms of the fusion center's classification accuracy, the network discriminant gain~\eqref{eq:Gtotal} provides a good proxy for it. In Section~\ref{sec:experiments}, several numerical experiments will showcase that policies achieving higher network discriminant gain also achieve superior classification performance.

\section{Goal-Oriented System Design}\label{sec:design}

The sensing model assumes the collaboration of $K$ devices, with device $k$ extracting $N_k$ features, yielding the total discriminant gain in \eqref{eq:Gtotal}. This gain depends on both physical and Medium Access Control (MAC) layer parameters:

\begin{itemize}
    \item \emph{At the physical layer,} the sensing power $P_{s,k}$ influences the application performance. Indeed, for the sensing application, the quality of the observation associated with each of the devices, encompassed in the probability distribution $p_{k,\ell}$ in \eqref{eq:prob_distribution}, affects the classification performance. While the distribution mean $\bm{\mu}_{k,\ell}$ and residual variance $\bm{\Sigma}_{k,\ell}$ are imposed by the sensing environment (\emph{e.g.} the device processing capacity, the physical location of the device, \dots) and cannot be controlled by the system, the sensing power is an optimization lever.
\item \emph{At the MAC layer,} because of quality of service and resource constraints, not all devices can be queried in a cycle. An important optimization lever is the scheduling policy of the fusion center, which selects the subset of devices to query at each cycle. The scheduling probability $\bm{\pi}$ should integrate the expected contribution of the device to the application goal, as in \eqref{eq:discriminant_gain}, and also its radio conditions, which define the cost of transmitting its data.

\end{itemize}

Our goal-oriented design framework aims at maximizing the application goal, modeled as the network discriminant gain~\eqref{eq:Gtotal} by acting on both the sensing power of the DFRC transceiver $\bm{P}_s = \left[P_{s,1}, \dots, P_{s,K}\right] \in {\mathbb{R}^{+}}^K$ and the scheduling policy of the network $\bm{\pi} \in [0,1]^K$. Hence, our application objective is
\begin{equation}
\label{eq:max_min_objective}
    \underset{\bm{\pi} \in [0,1]^K, \bm{P}_s \in {\mathbb{R}^{+}}^K}{\operatorname{maximize}} \;\; \min_{\ell\neq \ell^\prime} G^{(\ell,\ell^\prime)}(\bm{\pi},\bm{P}_s).
\end{equation}
This objective~\eqref{eq:max_min_objective} is maximized under different constraints reflecting the limited sensing capabilities of the devices, the network resources, the data service requirements, and the total energy. The modeling constraints are detailed below.
\subsection{Device power constraint}
The sensing power is limited by hardware specifications. Additionally, depending on the operating mode or battery level, the device can set a logical limit on the sensing power,
    \begin{equation}\label{eq:c1}
         P_{s,k} \leq P_{\max,k}, \quad k \in \{1,\cdots, K\}. \tag{C1}
    \end{equation}
\subsection{eMBB data rate constraint} We denote by $e_k$ the spectral efficiency of the $k$-th device, which can be computed from the Signal to Interference and Noise ratio (SINR) as $e_k = f(\text{SINR}_k)$ via the modified Shannon formula as in~\cite{jovanovic2014qos}, or using simulation-based link level curves as in~\cite{landre2013lte}. 
Assuming that the base station deploys a fair bandwidth-sharing strategy for regular data, the average rates of the three stages are as follows. During the stall stage, the $k$-device can transmit at the rate
    \begin{equation}
        r^{w}_k =  \frac{e_k B}{K},
    \end{equation}
    where $B$ is the communication bandwidth.  
 During the sensing stage, the $k$-device can only transmit if it has been scheduled for sensing by the fusion center. Hence, the $k$-th device transmits eMBB data with probability $1-\pi_k$. The $k$-th device's average transmission rate during the sensing stage is
    \begin{equation}
    \label{eq:sensing_rate}
        r^s_k =  (1 - \pi_k) \mathbb{E}\left[ \frac{e_k B}{1 + m_k} \right],
    \end{equation}
    where $m_k$ is the number of other devices sending eMBB data simultaneously with the $k$-th device. The eMBB data rate during the stall stage is simply $r^{w}_k = e_k B/K$. 
    During the feature transmission stage, the devices report the acquired features sequentially using the whole spectrum, and there is no eMBB transmission, \emph{i.e.}, $r^f_k = 0$. However, we have an additional constraint on the policy since $T_f$ is limited
    \begin{equation}
        \label{eq:c2.a}
        T_f \geq \sum^K_{k=1}\ \pi_k \frac{l_k}{e_k B}, \tag{C2.a}
    \end{equation}
    where $l_k$ is the length of the features. The constraint \eqref{eq:c2.a} discourages scheduling devices with poor radio conditions with high probability. We note that $T_f$ should be set to the maximum resource the system can afford to sacrifice for feature transmission, and after solving for a policy, the actual operating $T_f$ can be set to the RHS of~\eqref{eq:c2.a}.
    
    The final average eMBB rate for device $k$ is
    \begin{align} \label{eq:r_avg}
        r^{avg}_k &= \frac{1}{T} \left(T_w r^{w}_k + T_s r^s_k \right)   \nonumber \\
        &=\frac{e_k B}{T}\left(\frac{T_w}{K}+ T_s (1 - \pi_k)  \mathbb{E}\left[ \frac{1}{1 + m_k} \right] \right).
    \end{align}
    The base station should guarantee a minimum eMBB rate for each device, depending on its radio conditions:
    \begin{equation}\label{eq:c2.b}
        r^{avg}_k \geq r^{\min}_{k}, \quad k\in\{1,\cdots,K\}. \tag{C2.b}
    \end{equation}
    The guarantee levels $r^{\min}_k$ in constraint~\eqref{eq:c2.b} can be device-dependent, as devices with less favorable radio conditions consume more resources if guaranteed at the same throughput as devices with a large SINR. However, our model remains general and allows an arbitrary guarantee policy $[r^{\min}_{1}, \dots, r^{\min}_{K}]$. We show in Section~\ref{sec:experiments} how to select the guarantee level.
\subsection{Total energy constraint} The energy for the sensing service is bounded by the budget $E$:
    \begin{equation}\label{eq:c3}
        \sum^K_{k=1} \left( P_{s,k} T_s + P_{f,k} T_f \right) \pi_{k} \leq E, \tag{C3}
    \end{equation}
   $P_{f,k}$ being the feature transmission power of the $k$-th device.
  
\subsection{Constrained problem formulation}
The objective definition~\eqref{eq:max_min_objective} is elusive as the program is computationally intractable due to the combinatorial aspect of the objective function~\eqref{eq:Gtotal}. We propose instead maximizing the sum of pairwise network discriminant gains. The optimization of sum gain and worst-case gain only diverges from one another in extreme cases where the characteristics of device sensors are highly different, and sensor class distinguishability is highly fluctuating. With the above constraints, the \emph{optimal policy} for our goal-oriented scheduling problem reads
\begin{align}\label{eq:P0}
\underset{\bm{\pi}, \bm{P}_s}{\operatorname{maximize}} & \;\; G(\bpi, \bP_{s}) = \sum^{L}_{\ell=1} \sum^{L}_{\ell^\prime > \ell} G^{(\ell,\ell^\prime)}(\bpi, \bP_{s}) \nonumber \\
\text{subject to} &\;\;P_{s,k} \in \bbR^+,\, \pi_k \in [0,1], \quad k \in \{1,\dots,K\}, \nonumber \\
&\;\; \eqref{eq:c1} \text{  and  } \eqref{eq:c2.a}, \eqref{eq:c2.b} \text{  and  } \eqref{eq:c3} \tag{P0}.
\end{align}

\section{Independent scheduling policy}
\label{sec:independent_policy}

We here consider the simplest case where devices are scheduled independently, with policy $\bpi$, and $\pi_k$ is the probability that the $k$-th device is scheduled to perform sensing in a cycle. Given that the schedules are drawn independently, the aggregation of device gains at the fusion center is just a linear combination weighted by the scheduling policy $\bm{\pi}$, that is
\begin{equation}
    G^{(\ell,\ell^\prime)}(\bm{\pi},\bm{P}_s) = \sum^{K}_{k=1}  \pi_k G^{(\ell,\ell^\prime)}_k(P_{s,k}),
\end{equation}
The objective of \eqref{eq:P0} becomes
\begin{equation} \label{eq:G_total_ind}
   G(\bm{\pi},\bm{P}_s) = \sum^L_{l=1} \sum_{l^\prime > l} \sum^{K}_{k=1} \sum_{n_k=1}^{N_k}  \frac{(\mu_{n_k,\ell} - \mu_{n_k,\ell^\prime})^2}{\sigma^2_{n_k} + \eta^2_{n_k} / P_{s,k}},
\end{equation}
which is a concave function of $P_{s,k}$. In the following, we relax constraints of \eqref{eq:P0} to derive a tractable problem.

\subsection{Lower bound on eMBB data rate}

The eMBB data rate constraint \eqref{eq:c2.b} depends on the average rate during the sensing stage $r^s_k$, which involves an expectation of $1/(1+m_k)$. Let $b_k \in \{0,1\}$ be the Bernoulli schedule variable corresponding to the scheduling policy $\pi_k$, \emph{i.e.} $E[b_k = 1] = \pi_k$. One has
\begin{equation}
\label{eq:def_m_k}
    m_k = \bm{1}\{b_k=0\} \sum^K_{k^\prime \neq k} (1 - b_{k^\prime}).
\end{equation}
Assuming the scheduling policies are independent between devices, the probability mass function (PMF) of the $m_k$ is $P[m_k = m]= \sum_{A \in F_m} \prod_{i \in A} (1-\pi_i) \prod_{j \in A^c} \pi_j$, $F_m$ is the set of all subsets of $m$ integers selecting from $\{1,\dots,K\} \setminus \{k\}$, and the complement $A^c =\{1,\dots,K\} \setminus \{k\} \setminus A$.
Due to the combinatorial nature of the PMF, no closed-form formula is available for the expectation in~\eqref{eq:sensing_rate}. 

Since $1/(1+m_k)$ is a convex function in $m_k$, a lower bound on the sensing rate can be derived using Jensen's inequality:
\begin{equation} \label{eq:sensing_rate_lb_independent}
    r^s_k \geq (1 - \pi_k) \frac{e_k B}{1 + E[m_k]}
\end{equation}
Furthermore, from \eqref{eq:def_m_k}, one has
\begin{equation}\label{eq:expect_mk}
    \mathbb{E}[m_k]= K - 1 - \sum^K_{k^\prime \neq k} \pi_{k^\prime},
\end{equation}
The proof of~\eqref{eq:expect_mk} is deferred to the Appendix. Substituting~\eqref{eq:expect_mk} into \eqref{eq:sensing_rate_lb_independent} yields
\begin{equation}
    r^s_k \geq (1 - \pi_k) \frac{e_k B}{K - \sum^K_{k^\prime \neq k} \pi_{k^\prime}}.
\end{equation}
The lower bound on the eMBB data rate reads
\begin{equation}\label{eq:r_taylor}
    r^{avg}_k \geq \frac{e_k B}{T}\left(\frac{T_w}{K}+ T_s \frac{1 - \pi_k}{K - \sum^K_{i \neq k}\pi_i}  \right),\quad \forall k\in\{1,\cdots,K\}
\end{equation}
With \eqref{eq:r_taylor}, the linearization of the constraint~\eqref{eq:c2.b} is
\begin{align}
     \pi_k - 1 + \left(\frac{r^{\min}_{k} T}{e_k B T_s} - \frac{T_w}{K T_s}\right) \left(K - \sum^K_{i \neq k}\pi_i \right) \leq 0. \tag{$\tilde{\text{C2.b}}$}
    \label{eq:c2_relax}
\end{align}

\subsection{Adaptive McCormick envelopes}

Constraint~\eqref{eq:c3} is bilinear, and the optimization variables $\bm{\pi}$ and $\bm{P}_s$ are bounded. We use McCormick envelopes \cite{mccormick1976_nonconvex} to relax the bilinear~\eqref{eq:c3} into a fixed set of linear constraints as:
\begin{subequations}\label{eq:mccormick}
\begin{align}
    \sum^K_{k=1} \left( P_{max,k} T_s + P_{f,k} T_f \right) \pi_{k} & \leq E\\
    \sum^K_{k=1} P_{s,k} T_{r,k} + P_{f,k} T_f \pi_{k} & \leq E.
\end{align}
\end{subequations}
The constraints~\eqref{eq:mccormick} ensure convergence of optimization algorithms; however, they severely limit the feasible search space of variables. We propose adaptively changing the McCormick envelopes during optimization to improve exploration. Let us consider sub-segments $[\bpi^{L}, \bpi^{U}]$ and $[\bP^{L}_s, \bP^{U}_s]$ of the ranges of $\bpi$ and $\bP_s$ respectively. The local McCormick envelopes are: 
\begin{align}
    \label{eq:c3a_relax} \sum^K_{k=1} \left( P^{U}_{s,k} T_s + P_{f,k} T_f \right) \pi_{k} + \left(P_{s,k} - P^{U}_{s,k} \right) T_s \pi^{L}_k & \leq E   \tag{$\tilde{\text{C3.a}}$}\\
    \label{eq:c3b_relax} \sum^K_{k=1} \left( P^{L}_{s,k} T_s + P_{f,k} T_f \right) \pi_{k} + \left( P_{s,k} - P^L_{s,k} \right) T_s \pi^{U}_k  & \leq E. \tag{$\tilde{\text{C3.b}}$}
\end{align}

\begin{figure}
    \centering
    \includegraphics[width=\linewidth, trim={0 5 0 20}, clip]{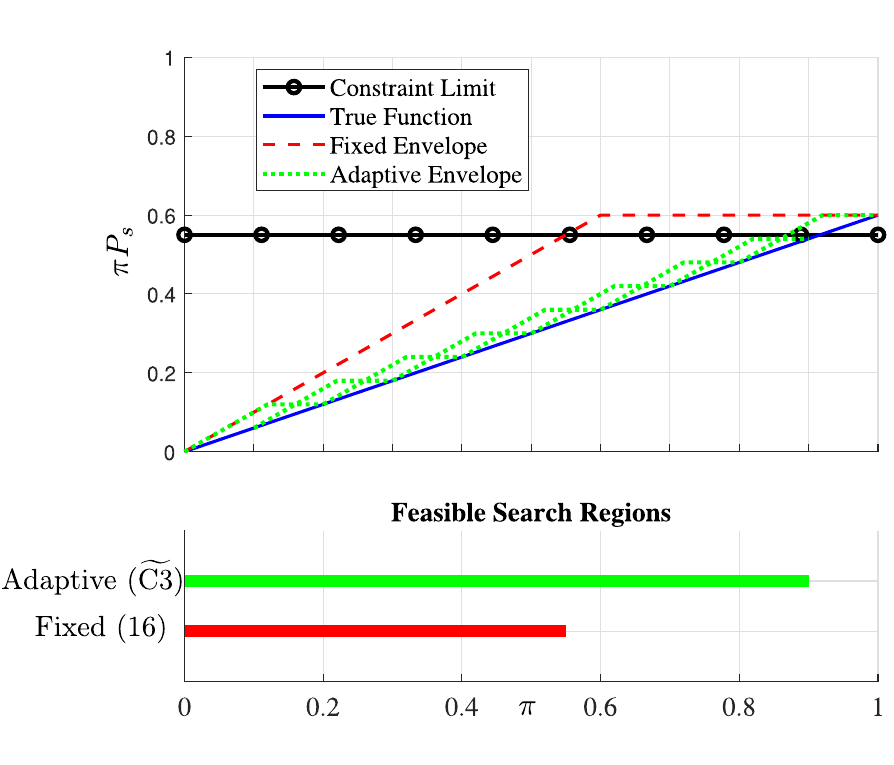}
    \caption{Feasible search regions of the adaptive and the fixed McCormick envelopes.}
    \label{fig:Adaptive_McCormick}
\end{figure}

As an example, the feasible search regions on $\bpi$ when $P_s=0.6$ are depicted in Figure~\ref{fig:Adaptive_McCormick}. One can see that the adaptive McCormick envelopes radically enlarge the feasible search region and thus allow better resource utilization.

\subsection{Relaxed independent scheduling problem}
With the relaxation \eqref{eq:c2_relax}, \eqref{eq:c3a_relax} and~\eqref{eq:c3b_relax}, the approximate optimal policy becomes
\begin{align}\label{eq:P1}
\underset{\bm{\pi}, \bm{P}_s}{\operatorname{maximize}} & \;\; G(\bm{\pi},\bm{P}_s) \nonumber \\
\text{subject to} & \;\;\pi_{k} \in [0,1], \quad k \in \{1,\dots,K\}, \nonumber \\
&\;\;P_{s,k} \in \bbR^+, \quad k \in \{1,\dots,K\}, \nonumber \\
&\;\; \eqref{eq:c1} \text{  and  } \eqref{eq:c2.a}, \eqref{eq:c2_relax} \text{  and  } \eqref{eq:c3a_relax}, \eqref{eq:c3b_relax}. \tag{P1}
\end{align}

While the constraints of~\eqref{eq:P1} are linear, the optimization program~\eqref{eq:P1} is not convex, as the objective~\eqref{eq:G_total_ind} to maximize is not a concave function. Yet, the program~\eqref{eq:P1} amounts to maximizing a sum of multiple fractional terms, often called the sum-of-ratios problem. In general, sum-of-ratios problems are NP-hard~\cite{Freund2001-sum_ratio}. Fortunately, given the convex constraints and the concave-over-convex form of the ratios, the simplified problem can be solved using the Jong method \cite{jong2012efficient}, which reparameterizes the problem as a parametric convex program. Moreover, it is to date the most efficient method for problems of this form~\cite{qian2024_sumratio}. The Jong method offers a theoretical guarantee of global optimality and a linear convergence rate \cite{jong2012efficient}. This is true if the fixed envelopes are deployed. For the adaptive envelopes, convergence is not guaranteed as the constraints change adaptively. However, we observe similar convergence rates for the two methods in experiments.

\section{Joint scheduling policy}
\label{sec:joint_policy}

To fully optimize for the common goal, we need to account for correlations among features across devices. Without loss of generality, we assume the same number of features $N_k$ for all devices. The joint conditional distribution is
\begin{equation}
    p_\ell = \cN \left(\bmu_\ell, \bGamma \right),
\end{equation}
where $\bm{\mu}_{\ell} = \left[ \bm{\mu}_{1,\ell}^T,\dots,\bm{\mu}_{K,\ell}^T \right]^T \in \mathbb{R}^{K N_k}$, and
\begin{equation*}
    \bm{\Gamma} = \begin{bmatrix} 
                        \bm{\Gamma}_{1,1} & \cdots &\bm{\Gamma}_{1,K} \\
						  \vdots &\ddots &\vdots \\
                        \bm{\Gamma}_{K,1} & \cdots & \bm{\Gamma}_{K,K} 
                    \end{bmatrix} \in \mathbb{R}^{K N_K \times K N_K},
\end{equation*}
 $\bGamma_{k,k} = \bSigma_k + \frac{1}{P_{s,k}} \diag(\bm{\eta}^2_{k}) $, and $\bGamma_{k,k^\prime} $ reflects the correlation between features from the $k$- and $k^\prime$-devices, which can change over time. 
The pairwise network joint discriminant gain is
\begin{align}
    \tilde{G}& ^{(\ell, \ell^\prime)} (\bpi, \bP_{s}) = \nonumber \\
    = &E \left[ \left( (\bb \otimes \bm{1}_{N_k}) \circ (\bm{\mu}_\ell - \bm{\mu}_{\ell^\prime}) \right)^T \bm{\Gamma}^{-1} \left( (\bb \otimes \bm{1}_{N_k}) \circ (\bm{\mu}_\ell - \bm{\mu}_{\ell^\prime}) \right) \right] \nonumber \\
     = &E \left[ \sum_i \sum_j b_{k_i} b_{k_j}  (\mu_{i,\ell} - \mu_{i,\ell^\prime}) (\mu_{j,\ell} - \mu_{j,\ell^\prime})   \bm{\Gamma}^{-1}_{i,j} \right] \nonumber \\
     = &\sum_i E[b_{k_i} b_{k_i}]  (\mu_{i,\ell} - \mu_{i,\ell^\prime})^2   \bm{\Gamma}^{-1}_{i,i}  + \nonumber \\
        &\sum_i  \sum_{j \neq i} E[b_{k_i} b_{k_j}] (\mu_{i,\ell} - \mu_{i,\ell^\prime}) (\mu_{j,\ell} - \mu_{j,\ell^\prime}) \bm{\Gamma}^{-1}_{i,j}, \label{eqn:pairwise_network_joint_gain}
\end{align}
where $\bb \in \{0,1\}^{K}$ is the binary scheduling variable, $\bm{1}_{N_k}$ is the vector of length $N_k$ with value one at all entries, and $k_i$ is the device index of the $i$-th feature. 

The $K$-variate Bernoulli distribution makes the joint scheduling depend on a $2^K$ parameters, i.e. with an exponential complexity w.r.t. the number of devices. Fortunately, the discriminant gain \eqref{eqn:pairwise_network_joint_gain} only depends on the first and second moments of the distribution. We can then optimize without dealing with an exponential number of parameters. 

To distinguish from the independent policy, we denote the joint scheduling parameters as $\bPi \in [0,1]^{K \times K}$, whose entries are moments of the joint distribution: $\Pi_{kk}=E[b_k^2] = E[b_k]$ and $\Pi_{k k^\prime} = E[b_k b_{k^\prime}]$. $\bPi$ is inherently a symmetric matrix, and thus has $K(K+1)/2$ degrees of freedom. Furthermore, its entries have to satisfy the Fréchet-Hoeffding bounds~\cite{Frechet1935}
\begin{equation} \label{eq:c4.a}
    \max(0, \Pi_{k, k} + \Pi_{k^\prime, k^\prime} - 1) \leq \Pi_{k, k^\prime} \leq \min (\Pi_{k,k}, \Pi_{k^\prime, k^\prime}), \tag{C4.a}
\end{equation}
while the corresponding covariance matrix should be positive semi-definite (PSD)
\begin{equation} \label{eq:c4.b}
    \bPi - \diag(\bPi) \diag(\bPi)^T \succeq 0. \tag{C4.b}
\end{equation}

Let $\brho \in \bbR^{K N_k \times K N_k}$ be the corresponding correlation matrix of $\bGamma$, one has the decomposition
\begin{align}
    \bGamma &= \bD \brho \bD,
\end{align}
where 
$\bD = \diag \left( \sqrt{\sigma^2_{1_1} + \frac{\sigma^2_r}{ P_{s,1}} }, \sqrt{\sigma^2_{2_1} + \frac{\sigma^2_r}{ P_{s,1}} }, \dots, \sqrt{\sigma^2_{n_K} + \frac{\sigma^2_r}{ P_{s,K}} } \right)$.

Assuming $\rho$ is invertible, the pairwise network joint discriminant gain can be written as
\begin{align}
    \tilde{G}^{(\ell, \ell^\prime)} (\bpi, \bP_{s}) &= \sum_i \frac{\Pi_{k_i, k_i}  (\mu_{i,\ell} - \mu_{i,\ell^\prime})^2}{D^2_{i,i}}   \brho^{-1}_{i,i}  +  \label{eqn:network_joint_gain_2}\\
        &\sum_i  \sum_{j \neq i} \frac{\Pi_{k_i, k_j} (\mu_{i,\ell} - \mu_{i,\ell^\prime}) (\mu_{j,\ell} - \mu_{j,\ell^\prime})}{D_{i} D_{j}} \brho^{-1}_{i,j} .\nonumber 
\end{align}

The first term in ~\eqref{eqn:network_joint_gain_2} is a sum of separate gains of features, and the second term accounts for cross-gains contributed by feature pairs due to correlation. The first term collapses into the network discriminant gain \eqref{eq:G_total_ind} when there is no correlation between the devices, \emph{i.e.} $\brho = \bm{I}$. Unfortunately, the additional cross-gains have a non-convex function of the sensing power in the denominator. This makes maximizing the exact network discriminant an NP-hard problem. Besides, it is not easy, if not infeasible, to obtain the feature-level correlation matrix $\brho$ in \eqref{eqn:network_joint_gain_2}. In the following, we propose a more practical joint-gain model that accounts for correlations at device level.

\subsection{Simplified Joint Gain Model}

We propose to use coefficients $c_{k, k^\prime}$ to capture the behavior of the joint discriminant gain under correlations as
\begin{align} \label{eq:approx_gain_corr}
    &\hat{G} (\bPi, \bP_{s}) = \sum^K_{k=1}  \left(\Pi_{kk} + \sum_{k^\prime \neq k} \Pi_{kk^\prime} c_{k, k^\prime} \right) G_k \nonumber \\
    &= \sum^K_{k=1}  \Pi_{kk} G_k + \sum^K_{k=1} \sum_{k^\prime > k} \Pi_{kk^\prime} c_{k, k^\prime} \left( G_k + G_{k^\prime} \right),
\end{align}
in which, with a slight abuse of notation, $G_k$ refers to 
\begin{equation}
    G_k(P_{s,k}) = \sum^L_{l=1} \sum_{l^\prime > l} \sum_{n_k=1}^{N_k}  \frac{(\mu_{n_k,\ell} - \mu_{n_k,\ell^\prime})^2}{\sigma^2_{n_k} + \eta^2_{n_k} / P_{s,k}}.
\end{equation}
We note that $c_{k, k^\prime} = c_{k^\prime, k}$, forming a symmetric matrix $\bc$.

The simplified joint gain model is a weighted sum of device gain, and it can be decomposed into two intuitive terms. The first term is exactly the network discriminant gain \eqref{eq:G_total_ind} of the independent scheduling policy. The second term represents the cross-gain effect. Particularly, it is the accumulation of the pairwise gain sum weighted by the joint scheduling probability $\Pi_{k, k^\prime}$ and $c_{k, k^\prime}$. One interpretation is that the coefficients measure the fraction of the original independent device gain that will be either added to or subtracted from the total gain when a pair of devices is co-scheduled. When the coefficients are zero, the simplified model is equivalent to the network discriminant gain of the independent scheduling policy.

Furthermore, the weighted-sum form preserves the structure of the objective function. More precisely, the simplified joint gain is a sum of concave-over-convex ratios and can thus be optimized efficiently by \cite{jong2012efficient}. Additionally, the model offers practicality for deployment in real-world scenarios, as the number of parameters capturing correlations is only $K(K-1)/2$, whereas that of the exact model is much larger, $KN_k(KN_k-1)/2$, since typically $N_k \gg K$. Unlike the statistics $\bmu, \bGamma_k, \bm{\eta}_{k}$, which can be obtained beforehand, the correlation-related parameters $\bc$ and $\brho$ depend on the relative position of devices and can vary over time. Hence, the enormous number of parameters needed to be estimated or measured makes the exact model inappropriate to deploy in the real world. Having said that, for theoretical analyses, the coefficients $\bc$ can be estimated from $\brho$ using simple linear regression. A scheme will be presented in Section~\ref{sec:joint_scheduling_experiments}.

\subsection{Relaxed joint scheduling problem}

To obtain a similar tractable lower bound on the eMBB rate, we have to derive the expected number of devices transmitting simultaneously with the $k$-th device (see the Appendix):
\begin{equation}
     E[m_k]	= \frac{1}{1 - \Pi_{kk}}  \sum^K_{k^\prime \neq k} \left[ 1 - (\Pi_{kk} + \Pi_{k^\prime k^\prime} - \Pi_{k k^\prime}) \right], \label{eq:E_m_k_joint}
\end{equation}
With \eqref{eq:E_m_k_joint}, the lower bound on eMBB data rate becomes

\begin{align}
    &\frac{e_k B}{T} \Bigg[ \frac{T_w}{K} + T_s (1 - \Pi_{kk}) \;\times \nonumber \\ 
    & \qquad \left(1 + \frac{1}{1 - \Pi_{kk}} \sum^K_{k^\prime \neq k} \left[ 1 - (\Pi_{kk} + \Pi_{k^\prime k^\prime} - \Pi_{k k^\prime}) \right] \right)^{-1} \Bigg] \geq r^{min}_k \nonumber \\
    &(1 - \Pi_{kk})^2 \geq \left( \frac{r^{min}_k T}{e_k B T_s} - \frac{T_w}{K T_s} \right) \times \nonumber \\
    & \hspace{8em} \left( K(1 - \Pi_{kk}) - \sum^K_{k^\prime \neq k} (\Pi_{k^\prime k^\prime} - \Pi_{k k^\prime}) \right)
\end{align}

The lower bound is more complicated and no longer presents a convex search space. Applying the first-order Taylor approximation to the convex LHS around some reference $\bnu$ gives a pessimistic linear constraint for the eMBB data rate.
\begin{align} \label{eq:c2_relax_joint}
    &(1 - \nu_{k})^2 - 2(1 - \nu_{k})(\Pi_{kk} - \nu_k) \geq \nonumber \\
    &\left( \frac{r^{min}_k T}{e_k B T_s} - \frac{T_w}{K T_s} \right) \left( K(1 - \Pi_{kk}) - \sum^K_{k^\prime \neq k} (\Pi_{k^\prime k^\prime} - \Pi_{k k^\prime}) \right). \tag{$\widehat{\text{C2.b}}$}
\end{align}

The relaxed joint scheduling problem becomes 
\begin{align}\label{eq:P2}
\underset{\bPi, \bm{P}_s}{\operatorname{maximize}} & \;\; \hat{G} (\bPi, \bP_{s}) \nonumber \\
\text{subject to} & \;\;\bPi \in [0,1]^{K \times K}, \nonumber \\
&\;\;P_{s,k} \in \bbR^+, \quad k \in \{1,\dots,K\}, \nonumber \\
&\;\; \eqref{eq:c1} \text{  and  } \eqref{eq:c2.a}, \eqref{eq:c2_relax_joint} \text{  and  } \eqref{eq:c3a_relax}, \eqref{eq:c3b_relax}, \nonumber \\
&\;\;   \text{  and  } \eqref{eq:c4.a}, \eqref{eq:c4.b}. \tag{P2}
\end{align}

Thanks to the sum-of-ratios structure of the simplified joint gain model, the program~\eqref{eq:P2} can be solved efficiently with the Jong method \cite{jong2012efficient} as all constraints are convex. We note that $\nu_k$ in~\eqref{eq:c2_relax_joint} can be set to the temporary solution of $\Pi_{kk}$ since iterative optimization is deployed.

\subsection{Joint sampling}

Unlike the independent scheduling policy, in which sampling for a schedule $\bb$ is straightforward given $\bpi$, sampling $\bb$ is no longer elementary since the scheduling parameters $\bPi$ describe the multivariate Bernoulli distribution only partially. We present in the following two procedures to sample $\bb$ whose moments match exactly or approximately $\bPi$.

\subsubsection{Ising Model}

The Ising model from statistical mechanics is a classical graphical model, in which each node is associated with a Bernoulli random variable. It is a simplified version of the multivariate Bernoulli since it considers only pairwise interactions. The Ising model is the unique maximum entropy distribution given the first and second moments of the variables \cite{Graphical_models_Wainwright_Jordan}. We note that our scheduling policy only designs the first and second moments of the $K$-variate Bernoulli distribution, making the Ising model is a suitable candidate to represent the scheduling policy $\bPi$. The likelihood of a state $\bb$ is 
\begin{equation}
    p(\bb) = \frac{1}{Z} \exp \left( \sum^K_{k=1} h_k b_k + \sum^K_{k=1} \sum^K_{k^\prime > k} J_{k,k^\prime} b_k b_{k^\prime}  \right),
\end{equation}
 $h_k$ models a node potential, $J_{k,k^\prime}$ denotes the interaction, and $Z(\bh, \bJ)$ the partition function which normalizes the density.

Given the convexity of the problem of solving for $\bh$ and $\bJ$ \cite{Graphical_models_Wainwright_Jordan}, one can fit an Ising model given a moment matrix $\bPi$ using Algorithm~\ref{alg:exact_ising_pi}, then sample $\bb$ from it. However, this is only possible when one can iterate over every possible schedule decision, \emph{i.e.}, the state space. The state space has a size of $2^K$; hence, the complexity increases exponentially with $K$. When iterating over the state space is prohibitive, one has to rely on Markov chain Monte Carlo (MCMC) methods to approximate gradients, as in \cite{Ackley1985}. After $\bh$ and $\bJ$ are obtained, one can calculate probabilities of all states and draw samples for small $K$. When $K$ is large, MCMC techniques can be used to sample from a high-dimensional Ising model. Nonetheless, the complexity may be computationally intractable due to exponential mixing time \cite{levin2017markov}, or the total cost of fitting and sampling the Ising model could exceed that of designing the policy itself. The next section introduces a scalable sampling technique that approximates given moments.

\begin{algorithm}
\caption{Gradient-based Ising model fitting}
\label{alg:exact_ising_pi}
\begin{algorithmic}[1]

\Require Target moment matrix $\boldsymbol{\Pi} \in \mathbb{R}^{K \times K}$
\Statex \quad \; Learning rate $\alpha$, tolerance $\epsilon$
\Ensure Parameters $\mathbf{h} \in \mathbb{R}^K$, $\mathbf{J} \in \mathbb{R}^{K \times K}$

\State \textbf{Initialize:} $\mathbf{h} \gets \mathbf{0}$, $\mathbf{J} \gets \mathbf{0}$
\State \textbf{Enumerate state space:} Generate $\mathbf{S} \in \{0,1\}^{2^K \times K}$ containing all binary vectors

\While{error $> \epsilon$}

    \State \textit{1. Compute probabilities for all states}
    \ForAll{state vectors $\mathbf{b} \in \mathbf{S}$}
        \State $E(\mathbf{b}) \gets \mathbf{h}^\top \mathbf{b} + \frac{1}{2}\mathbf{b}^\top \mathbf{J}\mathbf{b}$
    \EndFor
    \State $Z \gets \sum_{\mathbf{b} \in \mathbf{S}} \exp(E(\mathbf{b}))$
    \State $p(\mathbf{b}) \gets \frac{1}{Z}\exp(E(\mathbf{b})) \quad \forall \mathbf{b} \in \mathbf{S}$

    \State \textit{2. Compute model moments matrix}
    \State $\boldsymbol{\Pi}_{\text{model}} \gets \sum_{\mathbf{b} \in \mathbf{S}} p(\mathbf{b})\, \mathbf{b}\mathbf{b}^\top$

    \State \textit{3. Calculate gradient}
    \State $\boldsymbol{\Delta} \gets \boldsymbol{\Pi} - \boldsymbol{\Pi}_{\text{model}}$

    \State \textit{4. Update parameters}
    \State $\mathbf{h} \gets \mathbf{h} + \alpha\, \text{diag}(\boldsymbol{\Delta})$
    \State $\mathbf{J} \gets \mathbf{J} + \alpha\, \boldsymbol{\Delta}$
    \For{$k = 1$ to $K$}
        \State $\mathbf{J}_{kk} \gets 0$ \Comment{Zero out the diagonal}
    \EndFor

    \State error $\gets \max_{i,j} \lvert \boldsymbol{\Delta}_{ij} \rvert$

\EndWhile

\Statex \Return $\mathbf{h}, \mathbf{J}$

\end{algorithmic}
\end{algorithm}

\subsubsection{Dichotomized Gaussian}

The dichotomized Gaussian (DG) model was proposed in \cite{Macke_DG_spike} to generate binary random vectors with specified correlations. More precisely, the authors proposed a framework to generate synthetic spike trains by dichotomizing (binarizing) a multivariate Gaussian random variable. The idea is to sample from a multivariate Gaussian distribution, then use a threshold to obtain the desired Bernoulli sample. If $\bz \sim \mathcal{N}(\bm{0}, \bSigma^{DG})$ denotes the latent Gaussian random vector, a sample of the DG $\bb$ would be
\begin{equation}
    \bb = [ \mathbb{I}(z_k \leq \tau_k)]^K_{k=1},
\end{equation}
where $\tau_k =  \Phi^{-1}(\Pi_{kk})$ is the threshold corresponding to the $k$-th device assuming that every $z_k$ has a unit variance, $\Phi^{-1}$ is the inverse CDF of the standard normal distribution.

The main problem here is to determine the non-diagonal entries of $\bSigma^{DG}$ as the diagonal entries $\Sigma^{DG}_{k k}$ are equal to one. For every pair, we have to solve
\begin{equation}
    \label{eq:DG_correlation}
    \Pi_{k k^\prime} = \int_{-\infty}^{\tau_k} \int_{- \infty}^{\tau_k^\prime} \phi_2(z_k,z_{k^\prime};\Sigma^{DG}_{k k^\prime}) d z_{k^\prime} d z_k,
\end{equation}
where $\phi_2$ is the PDF of the standard bivariate normal distribution. 
Although the integral in~\eqref{eq:DG_correlation} has no analytical formula, we can solve~\eqref{eq:DG_correlation} efficiently using numerical computation since the function is monotonic in $\Sigma^{DG}_{k k^\prime}$. After $\btau$ and $\bSigma^{DG}$ are obtained, sampling from the DG distribution is as convenient as sampling from a multivariate normal distribution.

Furthermore, experiments in \cite{Macke_DG_spike} demonstrated that the entropy of the DG model is close to the theoretical maximum Ising model for a wide range of moments. A drawback of the DG model is that it cannot capture all possible correlations of the multivariate Bernoulli. In particular, even though we can always find valid solutions $\Sigma^{DG}_{k k^\prime}$ for every pairwise correlation from \eqref{eq:DG_correlation}, the constructed $\bSigma^{DG}$ can be non-positive definite; hence, it is not a valid covariance matrix for the DG distribution. In that case, we need to project $\Sigma^{DG}_{k k^\prime}$ to its closest positive definite matrix before generating samples, and then the moments of the generated $\bb$ will match $\bPi$ approximately.

\section{Numerical Experiments}
\label{sec:experiments}
We implement the optimization program~\eqref{eq:P1} and~\eqref{eq:P2} in \textsc{Matlab} using the \textsc{CVX} software for disciplined convex programming~\cite{grant2014cvx}.
Two experiments are conducted to evaluate each of the proposed polices. For independent scheduling, multiple analyses are done on a synthetic Gaussian mixture dataset; then, on a FMCW radar simulation dataset generated using \textsc{MATLAB} radar toolbox. For joint scheduling, the simplified joint gain model is first compared against the exact joint gain model using theoretical data. Finally, the framework is applied to a Vehicle-to-Everything (V2X) scenario of cooperative 4D-radar fusion for object detection. The approximate optimal policy~\eqref{eq:P1} and~\eqref{eq:P2} is referred to as the \emph{optimal subset} and is compared against the \emph{fair-sensing} and \emph{importance-aware} policies. The fair sensing policy balances the sensing task for all devices, \emph{i.e.} $\pi_1 = \pi_2 = \dots = \pi_{K}$. The importance-aware scheme prioritizes devices with higher discriminant gains by setting $\pi_k = 1$ and allocating the highest possible power under the energy budget to the most discriminative devices, provided the constraints are satisfied. When the energy budget is reached, the scheduling policy for the remaining less discriminative devices is $\pi_k = 0$. The importance-aware principle is inspired by~\cite{Lan_2023_ProgressiveFTX}.

For all experiments, we set the communication bandwidth $B = 10$ MHz around the carrier frequency $3.5$ GHz, the durations of stages are $T_s = 2$ s, $T_f = 0.5$ s, and $T_w = 4$~s.
The eMBB throughput guarantees~\eqref{eq:c2.b} are selected by letting $\text{SINR}_k$ be the SINR level of the $k$-th device, which is quantized $\text{SINR}^{std}_k$ using a grid of $\{ -5, 0,..., 35 \}$ dB. Without sensing, the standard guarantee rate is $r^{std}_k = e^{std}_k B/K$, where the spectral efficiency $e_k^{std}$ is computed from $\text{SINR}^{std}_k$. The guarantee rate is computed as
\begin{equation}
    r^{min}_k = \gamma r^{std}_k, \quad k\in\{1,\cdots,K\},
\end{equation}
where $\gamma$ represents the \emph{guarantee level} common to all devices. We note that the guarantee level~$\gamma$ can be greater than 1, as it is a ratio w.r.t. the quantized rate $r^{std}_k$.

\subsection{Independent scheduling experiments}

\subsubsection{Classification on a Synthetic Gaussian Dataset}
\label{sec:synthetic_data}

In this experiment, we consider $L=2$ classes, an ensemble of $K=20$ devices, and $N = 10$ features per device. It is further assumed $\bm{\Sigma}_k = \bm{I}_{N}$ and $\bm{\eta}_{k}^2 = \bm{1}$.
The devices are located randomly in a crown of inner radius $100$ m and outer radius $1000$ m centered at the fusion center. The transmitted power is subject to the path loss and shadowing, leading to the received power:
\begin{equation}
    % P_{rx} = P_{tx} + G - L - (PL_{ref} + 10\alpha \log_{10} (d) + \psi),
    P_{rx} = P_{tx} - \beta-10\alpha \log_{10} (d) - \psi,
\end{equation}
where $\alpha$ is the path loss exponent, $d$ is the distance between the sender and receiver, $\beta$ is the constant path loss component and $\psi \sim \cN (0, \sigma^2_\psi)$ is the shadowing. The SINR is then computed by introducing a noise of spectral density $-174$ dBm/Hz, and interference is modeled by a noise rise of $6$dB. 
We consider five categories of devices with transmitting powers of $0.1, 0.2, 0.4, 0.6,$ and $1$ W. The maximum allowed sensing power $P_{\max,k}$ is set equal to the device's transmitting power.

\begin{figure}[t]
        \includegraphics[width=\linewidth, trim={10 90 10 100}, clip]{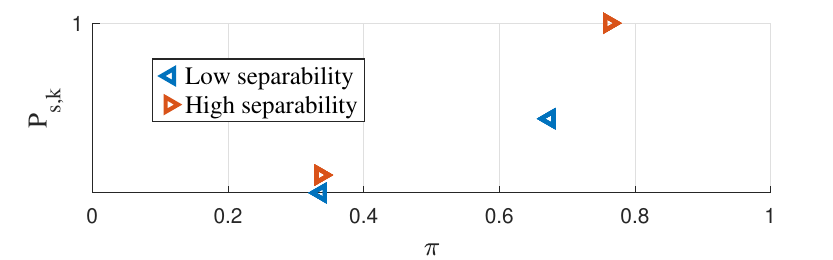}
        \caption{Illustration of optimal policy~\eqref{eq:P1} for two different separation between the class centers.
        In the simulation setting, each marker corresponds to 5 devices.}
    \label{fig:opt_policy_intuition}
\end{figure}

We start the simulation by assessing the effect of the separability between the classes on the optimal policy~\eqref{eq:P1}. Herein, 20 devices are divided into two groups depending on the level of separability, defined as the distance between the class centers $\mu_{k,1} - \mu_{k,2}$.
Figure \ref{fig:opt_policy_intuition} pictures the policies applied to the devices as a function of the separability perceived by the sensors. It can be appreciated that the optimal policy prioritizes devices with high separability. We note that some devices, which have $\pi_k > 0$ and $P_{s,k} = 0$, do not sense or send eMBB data during the sensing stage in order to save energy and dedicate bandwidth to others.

\begin{figure}
    \includegraphics[width=\linewidth, trim={20 48 10 50}, clip]{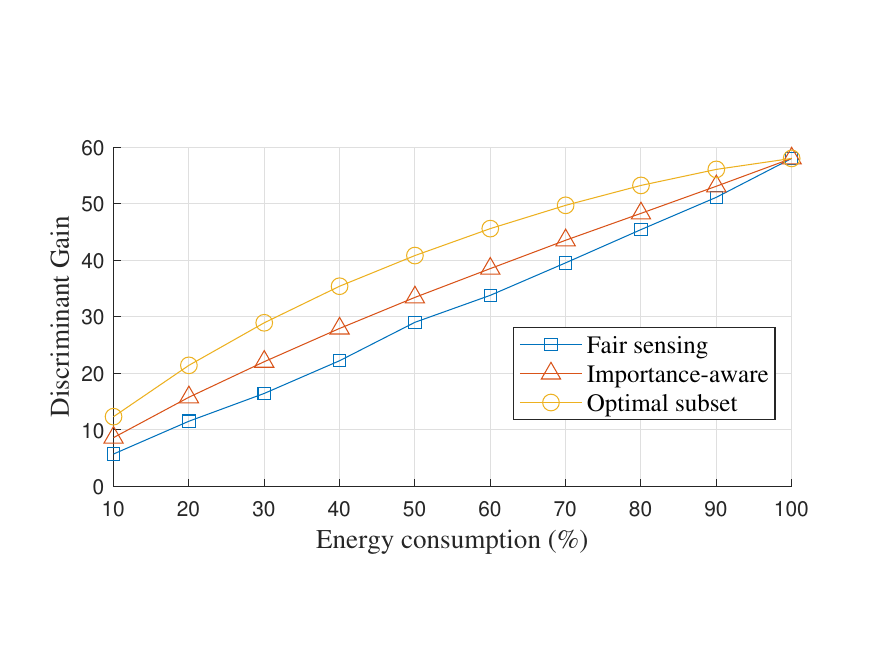}
        \caption{Discriminant gain vs. energy consumption level.}
    \label{fig:synthetic_data_energy}
\end{figure}

We now move to a more elaborate simulation setting, in which devices are randomly distributed within the considered zone. The classes are drawn such that $(\mu_{k,1} - \mu_{k,2}) \sim \cN(0,\bm{I}_{N})$. Letting the rate constraint be loose, $\gamma = 0.5$, the performance of the policies versus different levels of energy consumption is given in Figure~\ref{fig:synthetic_data_energy}. The energy consumption is shown in percentage w.r.t. the maximum value, \emph{i.e.}, the consumption when all devices are scheduled to sense at maximum power in every cycle. The optimal subset policy always yields the highest discriminant gain, while the importance-aware policy consistently outperforms the fair sensing policy. Only when it is allowed to schedule every device to sense in every cycle do the gains of the three policies coincide.

\begin{figure}
    \includegraphics[width=\linewidth, trim={20 48 10 50}, clip]{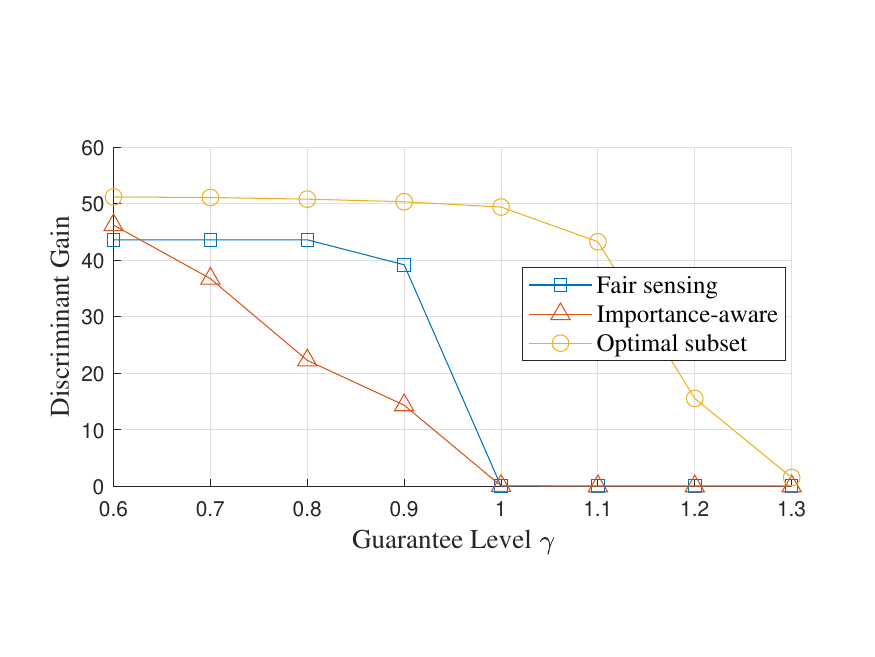}
        \caption{Discriminant gain vs. eMBB rate guarantee level $\gamma$.}
    \label{fig:synthetic_data_rate}
\end{figure}

In Figure~\ref{fig:synthetic_data_rate}, the eMBB rate guarantee level $\gamma$ varies while the energy constraint is set to 80\% of the maximum value. The gap between the discriminant gain of the optimal policy and the other two increases with higher values of the data service guarantee level~$\gamma$ until $\gamma\simeq 1.1$. The gain rapidly decreases above that threshold as sensing is impossible without sacrificing the mean eMBB performance.

Regarding the baselines, note that the importance-aware policy performs better when the energy constraint is activated, as it controls power allocation. Meanwhile, fair sensing is favorable when the eMBB rate constraint is tighter, as it optimizes the sensing probability. The results show the sensing benefit of our scheduling policy compared to the fair sensing and importance-aware policies under random statistics and various constraint conditions.

\subsubsection{Classification of Human Motions from Radar Samples}

Inspired by \cite{Wen_2024_ISCC}, we adapt our framework to classify child walking, child pacing, adult walking, and adult pacing patterns from FMCW signals. Unlike \cite{Wen_2024_ISCC}, where a primitive-based autoregressive hybrid channel model (PBAH) was employed, we fully utilize the radar signal simulated by a \textsc{Matlab} toolbox at 4 carrier frequencies: $5$ GHz, $12$ GHz, $24$ GHz, and $79$ GHz, while the sensing bandwidth is $10$ MHz for all bands. The heights of children and adults are uniformly distributed in the range $[0.9\text{m}, 1.2\text{m}]$ and $[1.5\text{m}, 2\text{m}]$, respectively. The walking and pacing speeds are $0.5H$ m/s and $0.25H$ m/s, respectively, where $H$ is the human height. 
The heading of moving humans is uniformly distributed in the $[-180, 180]$ degree interval. The sensing area is 20 meters wide by 40 meters long in front of the sensing device. A dataset comprising 800 training and 200 testing samples is created for each band. The samples are distributed uniformly across the classes, while the fusion center trains a support vector machine (SVM) to classify the human motions sensed by the devices.

\begin{figure}
  \centering
  \begin{subfigure}{\linewidth}
    \includegraphics[width=\linewidth, trim={10 60 10 70}, clip]{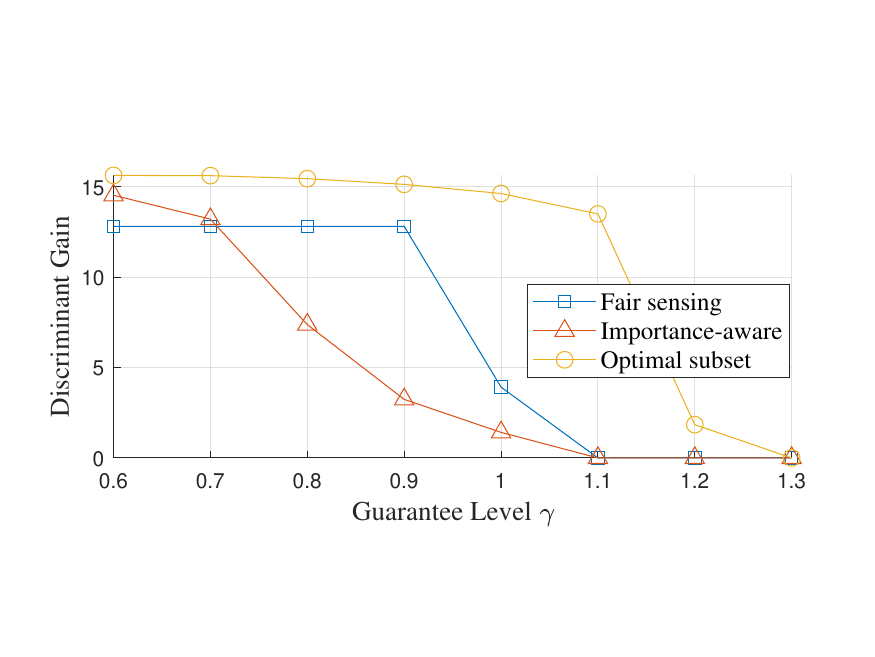}
    \caption{Theoretical discriminant gain against the guarantee level $\gamma$.}
        \label{fig:gain_radar}
  \end{subfigure}
  \begin{subfigure}{\linewidth}
    \includegraphics[width=\linewidth, trim={10 60 10 70}, clip]{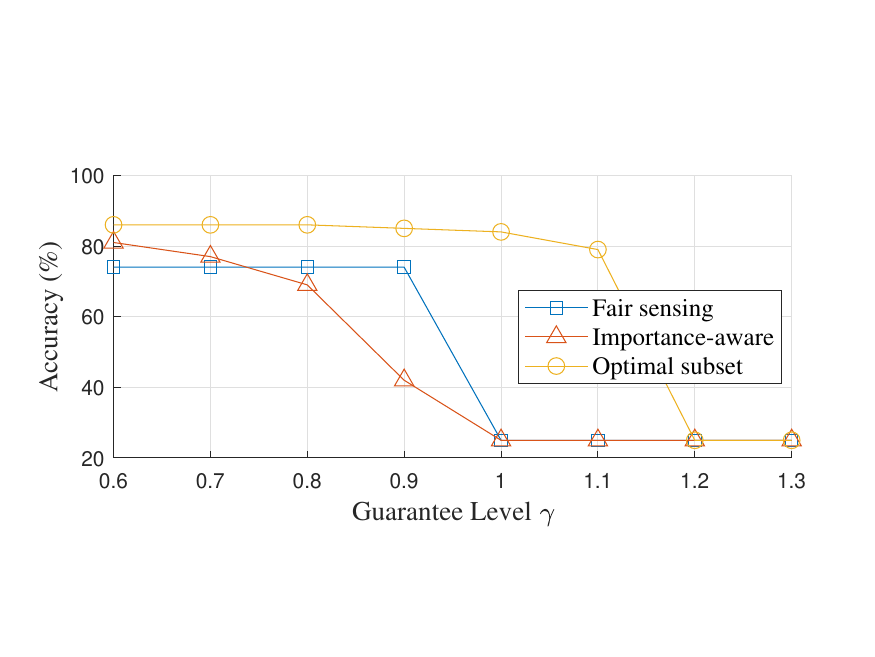}
    \caption{Accuracy of the fusion center classification with an SVM against the guarantee level $\gamma$.}
    \label{fig:accuracy_radar}
  \end{subfigure}
  \caption{Classification performance with MATLAB FMCW radar simulation data.}
  \label{fig:radar_simulation}
\end{figure}

We randomly distribute $K=20$ devices in $15$ regions and extract the $N_k=100$ leading PCA features of the device samples. We approximate the data distributions $p_{k,\ell}$ by Gaussians with the same sample means and covariances. We select a sensing power of $1$ W corresponding to an SNR of $20$ dB. The communication distance and energy constraints are the same as in the first experiment, and the sensing band is randomly selected from the 4 values.
Figure~\ref{fig:radar_simulation} shows the classification results.
It is interesting that the accuracy of SVM predictions follows the same trend as the theoretical gain, even though the network's classification objective relies on many modeling approximations. This suggests the robustness of the proposed scheme and its potential for practical application. When the discriminant gain falls to $0$, the fusion center accuracy is about 25\%, corresponding to that of a random guess.

\subsection{Joint scheduling experiments}
\label{sec:joint_scheduling_experiments}

\subsubsection{Simplified Joint Gain Model Validation on Synthetic Data}
We now validate the simplified joint gain model against the exact gain model. We consider $L=2$ classes, $K=3$ devices with $\bm{\Sigma}_k = \bm{I}_{N}$ and $\bm{\eta}_{k}^2 = \bm{1}$. We first select $N = 10$ features per device and compare the two models across different correlation levels, with a rate constraint $\gamma = 1.0$, while the energy constraint is removed. We generate the correlation matrix $\brho$ randomly keeping the cross-correlation magnitude $|\rho_{i,j}| \in [\rho_{\max}-0.1,\rho_{\max}]$. The devices are assumed to deploy a homogeneous feature extraction scheme, \emph{i.e.}, a feature can correlate with at most one feature of another device. These conditions make the inversion of $\brho$ more stable numerically. To obtain the coefficients $\bc$ of the simplified joint gain model, we propose a simple procedure to learn $\bc$ as follows.

Since $\bc$ is a symmetric matrix, and the diagonal entries have no meaning in our model, $K(K-1)/2$ parameters have to be estimated. We can solve a simple linear regression problem to fit $\bc$ to behave like cross-terms in~\eqref{eqn:network_joint_gain_2}. One can plug some $\bPi$ and $\bP_s$ to compute the cross-term gain in~\eqref{eqn:network_joint_gain_2} and expect the cross-term gain in the simplified model in~\eqref{eq:approx_gain_corr} to match this. By setting only a pair of devices active together $\Pi_{kk} = \Pi_{k^\prime k^\prime} = \Pi_{k k^\prime} = \Pi_{k^\prime k} = 1$ with $\bP_s = \bP_{\max}$, one has 
\begin{equation}
    \left(G_k(P_{\max,k}) + G_{k^\prime}(P_{\max,k^\prime}) \right) c_{k,k^\prime} = w,
\end{equation}
$w$ being the cross-term gains from the exact model~\eqref{eqn:network_joint_gain_2} given the same inputs, leading to the full $\bc$.
The fitted $\bc$ is then used as the input to solve for the optimal subset policy.

\begin{figure}
    \includegraphics[width=\linewidth, trim={20 40 10 50}, clip]{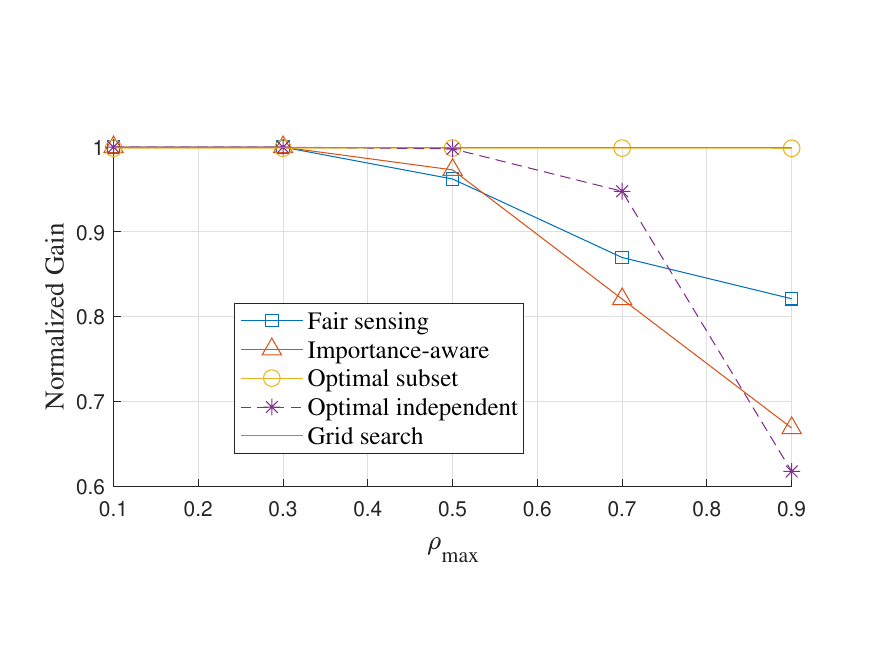}
        \caption{Joint discriminant gain against correlation level.}
    \label{fig:synthetic_corr_corr}
\end{figure}

Both the fair sensing and importance-aware policies work on the exact gain~\eqref{eqn:network_joint_gain_2} while the optimal subset is a solution to~\eqref{eq:P2} following the simplified joint gain. In addition, we implement a grid-search-based policy as a sub-optimal reference, as well as the optimal independent scheduling policy. The grid search is feasible only for a small number of devices ($K \leq 3$) and for grids with medium granularity. Here, the grids of the variables $\bPi$ and $\bP_s$ consist of 9 points equally spaced over their ranges. Figure~\ref{fig:synthetic_corr_corr} shows the joint discriminant gain against the correlation magnitude. In the weak-correlation regime, all policies yield equivalent performance. The optimal subset stands out when there are strong correlations. The optimal independent policy performs even worse than the fair and importance-aware policies for extremely high correlations.

\begin{figure}
     \includegraphics[width=\linewidth, trim={20 45 10 50}, clip]{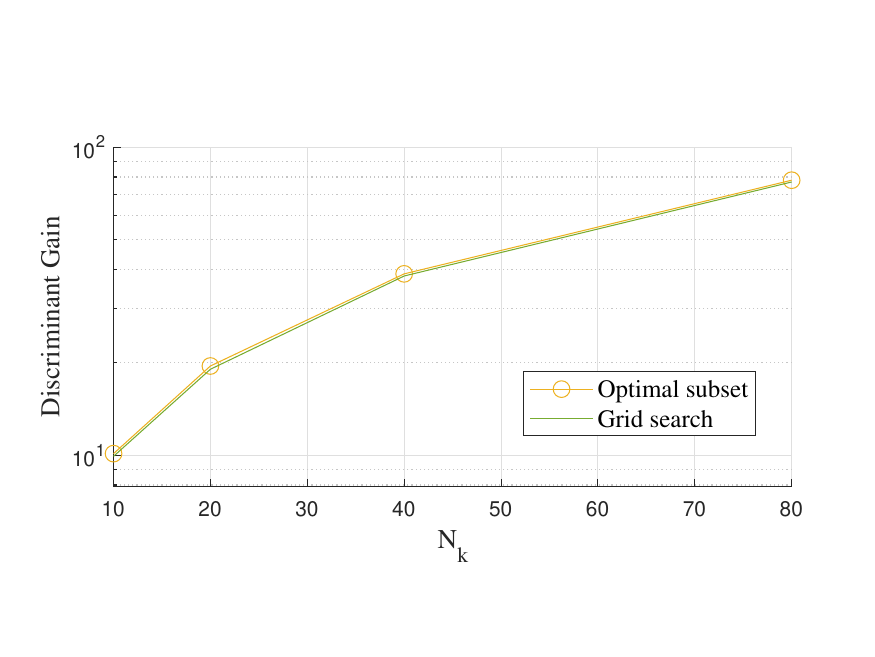}
        \caption{Simplified Gain Model performance versus $N_k$.}
    \label{fig:synthetic_corr_Nk}
\end{figure}

We further evaluate the simplified joint gain model as $N_k$ increases while the energy constraint is set to 80\%, $\gamma = 1$, and $|\rho_{ij}| \leq 0.5$. Since the exact model uses a $K N_k \times K N_k$ matrix to capture the correlation effect and the simplified model only uses a $K \times K$ matrix, it is reasonable to expect worse approximation as $N_k$ increases. Nevertheless, the results in Figure~\ref{fig:synthetic_corr_Nk} show that the simplified gain model approximation performance persists when $N_k$ increases, and it matches the performance of the sub-optimal grid search. It can be said that the simplified joint gain model is scalable with $N_k$.

\subsubsection{4D Radar Fusion for Object Detection}
\label{sec:4D_radar}

\begin{figure}
    \begin{subfigure}{0.49\linewidth}
        \includegraphics[width=1.2\linewidth, trim={35 0 75 0}, clip]{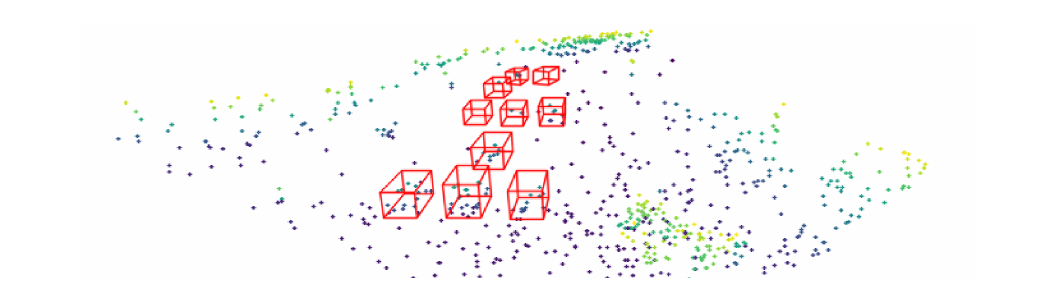}
        \caption{Device 1 data.}
    \end{subfigure}
    \begin{subfigure}{0.49\linewidth}
        \includegraphics[width=\linewidth, trim={155 20 15 0}, clip]{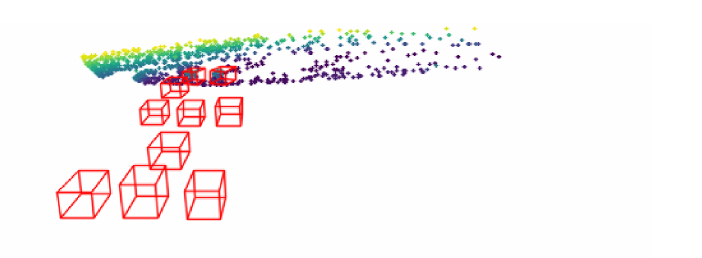}
        \caption{Device 2 data.}
    \end{subfigure}
    \begin{subfigure}{\linewidth}
        \includegraphics[width=\linewidth, trim={10 0 15 5}, clip]{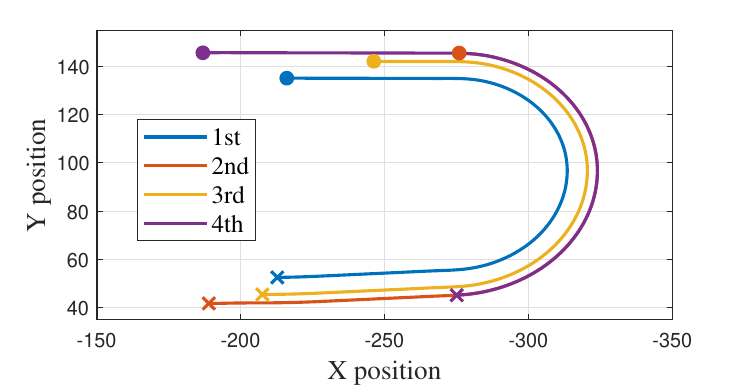}
        \caption{Trajectories in the scene \emph{2024\_06\_25\_15\_46\_09}, starting and ending positions are denoted by round and cross marks, respectively.}
    \end{subfigure}
    \caption{Visualization of the V2X-R dataset.}
    \label{fig:V2X-R-visualization}
\end{figure}

Our system model is particularly applicable to the collaborative object detection task in Vehicle-to-Everything (V2X) communication. Smart vehicles are often equipped with three main sensor types: LiDAR, cameras, and 4D radar. Our framework considers power allocation and scheduling; it is thus best suited to 4D radar, as there is typically no power control in LiDAR or camera systems. In addition, 4D radar offers many benefits compared to LiDAR and cameras, \emph{e.g.}, weather robustness and Doppler information \cite{Huang_2025_V2X-R}. We leave the problem of adaptation to other modalities for future work. The V2X-R dataset \cite{Huang_2025_V2X-R} is a recently published dataset containing LiDAR, camera, and 4D radar data for the cooperative 3D object detection task. The 4D radar data is stored in point cloud format, and a multi-view visualization from the dataset is shown in Figure~\ref{fig:V2X-R-visualization}. 

The authors of \cite{Huang_2025_V2X-R} proposed a cooperative fusion pipeline for 3D object detection and benchmarked it with multiple fusion strategies. We adapt their pipeline as follows. A goal-oriented policy determines which devices will perform sensing and how much power they will use. The scheduled devices comply with the policy and obtain raw point cloud data. They process the data locally using the PointPillar backbone \cite{Lang_2019_PointPillars} to obtain intermediate feature vectors for pillars, which they then send to the fusion center. The fusion center deploys AdaFusion \cite{Lai_AdaFusion} to fuse intermediate features and perform object detection for each pillar using a classification head. The chosen operating threshold of the classification head is $0.2$ \cite{Huang_2025_V2X-R}. The PointPillar backbone's idea is to divide space into pillars of the same size, \emph{e.g.}, 0.4 m $\times$ 0.4 m $\times$ 4 m. Then, the data points are grouped into pillars based on geometry, and feature extraction is performed intra-pillar and inter-pillar. As the V2X-R dataset was simulated without noise, we implemented our own noise simulation scheme.

The sensing power dictates the echo signal's SNR and thus determines the detection probability. We adopt a logistic function to approximate the S-shaped detection probability versus $\text{SNR}_\text{radar}$ curve induced by the Marcum Q-function \cite{luong2022_NoiseRadar_logistic},
\begin{equation}
    P_d(\text{SNR}_\text{radar}) = \frac{1}{1 + \exp(-\xi ( \text{SNR}_\text{radar} - \text{SNR}_0) )},
\end{equation}
where $\text{SNR}_0$ is the SNR level where the droprate is 50\%, $\xi$ denotes the smoothness of the droprate, and we set $\xi=0.3$ and $\text{SNR}_0=10$ dB. The SNR level of data points follows the practical $1/R^4$ backscattered power decay model \cite{richards2005fundamentals}
\begin{align}
    \text{SNR}_{\text{radar}}(R) &= \text{SNR}_\text{ref} - 40 \log_{10} \left( \frac{R}{R_\text{ref}} \right),
\end{align}
where $R$ is the distance from a data point to the radar transceiver, $\text{SNR}_\text{ref}$ and $R_\text{ref}$ are respectively the SNR and distance of a reference position.

The scene "2024\_06\_25\_15\_46\_09" \cite{Huang_2025_V2X-R} is chosen from the V2X-R dataset based on the criterion that multiple cooperating vehicles are not too far apart. The trajectories of vehicles in the scene are depicted in Figure~\ref{fig:V2X-R-visualization}. We downsampled the sampling rate from 20 to 4 frames per second. The 2024\_06\_25\_15\_46\_09 scene has a duration of 16.6 seconds, corresponding to 66 frames. The first 4 frames (1 second) are dedicated to estimating correlations between devices. This can be seen as a setup phase, in which all devices perform sensing and send features to the fusion center for correlation estimation. In practice, the setup phase occurs periodically as vehicles' relative positions change over time. We propose to estimate $\bc$ by observing changes in the prediction confidence. The prediction confidence is computed from the fused data of a pair of devices and compared with the mean confidence using data from the two devices separately. We only measure changes in the confidence of the positive class, \emph{i.e.}, when the classification head output exceeds the $0.2$ threshold. A normalization is done so that the worst pair coefficient is zero, and the final estimated coefficients are
\begin{equation*}
    \bc = \begin{bmatrix}
        0.000 & 0.858 & 0.654 & 0.000 \\
        0.858 & 0.000 & 0.848 & 0.876 \\
        0.654 & 0.848 & 0.000 & 0.681 \\
        0.000 & 0.876 & 0.681 & 0.000
    \end{bmatrix}.
\end{equation*}
This estimated gain indicates that combining data from devices 2 and 4 is better than combining data from devices 1 and 4. Indeed, the first-fourth pair is closer together, so their data are more correlated than those of the second-fourth pair.

\begin{figure}
    \centering
    \includegraphics[width=\linewidth,trim={25 5 25 50},clip]{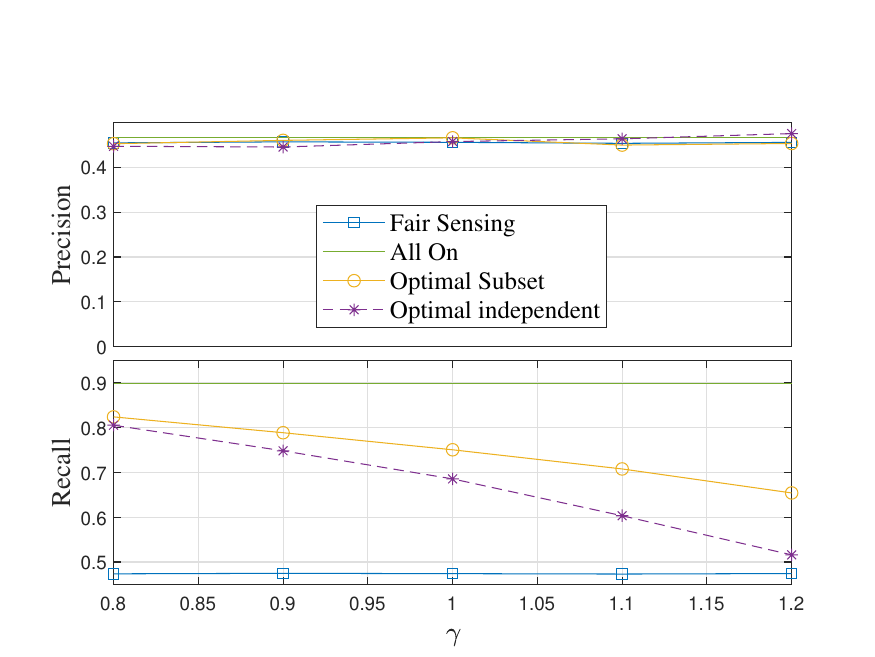}
    \caption{Policy performance evaluation on 4D radar data.}
    \label{fig:jan_20_eval}
\end{figure}

The performances of All On, Fair Sensing, and Optimal Subset and Optimal Independent policies are given in Figure~\ref{fig:jan_20_eval} under a 50\% maximum energy constraint, and $\gamma$ varies from 0.8 to 1.2. The importance-aware policy is omitted as it fails to produce a valid policy given the constraints. The Ising model is used for sampling. All On is an ideal policy for sensing, in which all devices always sense at their maximum power. The accuracy is not shown because it is not meaningful given the severe class imbalance. The operating threshold is $0.2$, so the precision is expected to be low even for the All On policy. There is no significant difference in precision between the policies. On the other hand, in recall, there is a large gap between the Optimal subset and the Fair sensing policy. The gap between the Optimal subset and the Optimal independent policies gradually increases as the communication constraint $\gamma$ becomes tighter. High values of $\gamma$ make it impossible to query all devices together; hence, the optimal choice of a subset of devices has a greater impact on the performance.

\section{Conclusion}
\label{sec:conclusion}

We proposed a scalable goal-oriented framework for collaborative ISAC that maximizes the performance of the classification task, measured through the discriminant gain, under eMBB throughput, resource consumption, device power, and energy constraints.
We derived two sub-optimal policies: independent scheduling and joint scheduling, whose parameters are the first and the first two moments of the sampling variable, respectively. The independent scheduling policy has low complexity, but misses fully collaborative control. The joint policy controls pairwise joint sampling probabilities and thus has higher complexity. We proposed a simplified joint-gain model to reduce design complexity. We formulated an associated constrained optimization problem and obtained the optimal policies through successive convex approximation. Comprehensive experiments with a perfect and imperfect Gaussian mixture prior demonstrated the benefits of the proposed scheme. 
Further study might consider extending the framework to other modalities, such as LiDAR and cameras.

% \bibliographystyle{IEEEtran}
% \bibliography{references} % Path to your BibTeX file
\renewcommand*{\bibfont}{\footnotesize}
\printbibliography

\appendix

\input{appendix}

\end{document}

%% file: header.tex
\usepackage{bm,bbm}
\usepackage{amsthm,amssymb,amsmath,mathtools}
\usepackage{stmaryrd}
\usepackage{babel,csquotes}
\makeatletter

\theoremstyle{plain}

\bmdefine{\bA}{A}
\bmdefine{\ba}{a}
\bmdefine{\bB}{B}
\bmdefine{\bb}{b}
\bmdefine{\bC}{C}
\bmdefine{\bc}{c}
\bmdefine{\bD}{D}
\bmdefine{\bd}{d}
\bmdefine{\bE}{E}
\bmdefine{\be}{e}
\bmdefine{\bF}{F}
\bmdefine{\bf}{f}
\bmdefine{\bG}{G}
\bmdefine{\bg}{g}
\bmdefine{\bH}{H}
\bmdefine{\bh}{h}
\bmdefine{\bI}{I}
\bmdefine{\bi}{i}
\bmdefine{\bJ}{J}
\bmdefine{\bj}{j}
\bmdefine{\bK}{K}
\bmdefine{\bk}{k}
\bmdefine{\bL}{L}
\bmdefine{\bl}{l}
\bmdefine{\bM}{M}
\bmdefine{\bmm}{m}
\bmdefine{\bN}{N}
\bmdefine{\bn}{n}
\bmdefine{\bO}{O}
\bmdefine{\bo}{o}
\bmdefine{\bP}{P}
\bmdefine{\bp}{p}
\bmdefine{\bQ}{Q}
\bmdefine{\bq}{q}
\bmdefine{\bR}{R}
\bmdefine{\br}{r}
\bmdefine{\bS}{S}
\bmdefine{\bs}{s}
\bmdefine{\bT}{T}
\bmdefine{\bt}{t}
\bmdefine{\bU}{U}
\bmdefine{\bu}{u}
\bmdefine{\bV}{V}
\bmdefine{\bv}{v}
\bmdefine{\bW}{W}
\bmdefine{\bw}{w}
\bmdefine{\bX}{X}
\bmdefine{\bx}{x}
\bmdefine{\bY}{Y}
\bmdefine{\by}{y}
\bmdefine{\bZ}{Z}
\bmdefine{\bz}{z}

\bmdefine{\balpha}{\alpha}
\bmdefine{\bbeta}{\beta}
\bmdefine{\bgamma}{\gamma}
\bmdefine{\bGamma}{\Gamma}
\bmdefine{\bdelta}{\delta}
\bmdefine{\bDelta}{\Delta}
\bmdefine{\bepsilon}{\epsilon}
\bmdefine{\bvarepsilon}{\varepsilon}
\bmdefine{\bzeta}{\zeta}
\bmdefine{\bmeta}{\eta}
\bmdefine{\btheta}{\theta}
\bmdefine{\bTheta}{\Theta}
\bmdefine{\biota}{\iota}
\bmdefine{\bkappa}{\kappa}
\bmdefine{\blambda}{\lambda}
\bmdefine{\bLambda}{\Lambda}
\bmdefine{\bmu}{\mu}
\bmdefine{\bnu}{\nu}
\bmdefine{\bpi}{\pi}
\bmdefine{\bPi}{\Pi}
\bmdefine{\brho}{\rho}
\bmdefine{\bsigma}{\sigma}
\bmdefine{\bSigma}{\Sigma}
\bmdefine{\btau}{\tau}
\bmdefine{\bupsilon}{\upsilon}
\bmdefine{\bUpsilon}{\Upsilon}
\bmdefine{\bphi}{\phi}
\bmdefine{\bPhi}{\Phi}
\bmdefine{\bchi}{\chi}
\bmdefine{\bpsi}{\psi}
\bmdefine{\bPsi}{\Psi}
\bmdefine{\bomega}{\omega}
\bmdefine{\bOmage}{\Omega}

\newcommand{\cN}{\mathcal{N}}

\newcommand{\bbR}{\mathbb{R}}

\DeclareMathOperator{\diag}{diag}

\let\hat\widehat
\let\tilde\widetilde

\bmdefine{\bWW}{WW}

\makeatother

%% file: appendix.tex
\section*{Derivations of the expected number of eMBB co-active devices}
\label{sec:appendix}
\textbf{Independent scheduling policy:}
\begin{align}
    \mathbb{E}[m_k] &= E \left[ \sum^K_{k^\prime \neq k} (1 - b_{k^\prime}) | b_k=0\right] \nonumber\\
                    & = E \left[ \sum^K_{k^\prime \neq k} (1 - b_{k^\prime})\right]  = \sum^K_{k^\prime \neq k} E \left[ 1 - b_{k^\prime}\right] \nonumber \\
                    & = K - 1 - \sum^K_{k^\prime \neq k} \pi_{k^\prime}.
\end{align}

\textbf{Joint scheduling policy:}

\begin{align}
     E[m_k]	&= E\left[ \sum^K_{k^\prime \neq k} (1 - b_{k^\prime}) \; \bigg| \; b_k = 0 \right]= \sum^K_{k^\prime \neq k} Pr[ b_{k^\prime} = 0 \; | \; b_k=0]  \nonumber \\
		%&= \sum^K_{k^\prime \neq k} Pr[ b_{k^\prime} = 0 \; | \; b_k=0]  \nonumber \\
		&= \frac{1}{Pr[b_k = 0]} \sum^K_{k^\prime \neq k} Pr[ b_{k^\prime} = 0, b_k=0] \nonumber \\
		&= \frac{1}{1 - \Pi_{kk}}  \sum^K_{k^\prime \neq k} \left( 1 - Pr[ b_{k^\prime} = 1 \cup b_k=1] \right) \nonumber \\
		&= \frac{1}{1 - \Pi_{kk}}  \sum^K_{k^\prime \neq k} \big( 1 - (Pr[ b_{k^\prime} = 1] + Pr[b_k=1] - \nonumber \\ & \phantom{=...........} Pr[ b_{k^\prime} = 1, b_k=1] \big) \nonumber \\
		&= \frac{1}{1 - \Pi_{kk}}  \sum^K_{k^\prime \neq k} \left[ 1 - (\Pi_{kk} + \Pi_{k^\prime k^\prime} - \Pi_{k k^\prime}) \right].
\end{align}